\documentclass[aps,prfluids,onecolumn,superscriptaddress]{revtex4-2}
\usepackage{bm}
\usepackage{comment}
\usepackage{amsmath}
\usepackage{amssymb}
\usepackage{enumitem}
\usepackage{hyperref}
\hypersetup{colorlinks=true,
allcolors=blue,}
\usepackage{graphicx}
\usepackage[dvipsnames]{xcolor}
\usepackage[normalem]{ulem}

\begin{document}

% Use the \preprint command to place your local institutional report
% number in the upper righthand corner of the title page in preprint mode.
% Multiple \preprint commands are allowed.
% Use the 'preprintnumbers' class option to override journal defaults
% to display numbers if necessary
%\preprint{}

%Title of paper
\title{Phenomenology of laminar acoustic streaming jets}

% repeat the \author .. \affiliation  etc. as needed
% \email, \thanks, \homepage, \altaffiliation all apply to the current
% author. Explanatory text should go in the []'s, actual e-mail
% address or url should go in the {}'s for \email and \homepage.
% Please use the appropriate macro foreach each type of information

% \affiliation command applies to all authors since the last
% \affiliation command. The \affiliation command should follow the
% other information
% \affiliation can be followed by \email, \homepage, \thanks as well.
\author{Bjarne Vincent}
\affiliation{INSA~Lyon, CNRS, Ecole Centrale de Lyon, Universite Claude Bernard Lyon~1, Laboratoire de Mécanique des Fluides et d'Acoustique, UMR5509, 69621, Villeurbanne France}
\affiliation{Fluid and Complex Systems Research Centre, Coventry University, Coventry CV15FB, United Kingdom}
\author{Daniel Henry}
\affiliation{INSA~Lyon, CNRS, Ecole Centrale de Lyon, Universite Claude Bernard Lyon~1, Laboratoire de Mécanique des Fluides et d'Acoustique, UMR5509, 69621, Villeurbanne France}
\author{Abhishek Kumar}
\affiliation{Fluid and Complex Systems Research Centre, Coventry University, Coventry CV15FB, United Kingdom}
\author{Valéry Botton}\email[]{valery.botton@insa-lyon.fr}
%\homepage[]{Your web page}
%\thanks{}
%\altaffiliation{}
\affiliation{INSA~Lyon, CNRS, Ecole Centrale de Lyon, Universite Claude Bernard Lyon~1, Laboratoire de Mécanique des Fluides et d'Acoustique, UMR5509, 69621, Villeurbanne France}
\author{Alban Pothérat}
\affiliation{Fluid and Complex Systems Research Centre, Coventry University, Coventry CV15FB, United Kingdom}
\author{Sophie Miralles}\email[]{sophie.miralles@insa-lyon.fr}
\affiliation{INSA~Lyon, CNRS, Ecole Centrale de Lyon, Universite Claude Bernard Lyon~1, Laboratoire de Mécanique des Fluides et d'Acoustique, UMR5509, 69621, Villeurbanne France}

%Collaboration name if desired (requires use of superscriptaddress
%option in \documentclass). \noaffiliation is required (may also be
%used with the \author command).
%\collaboration can be followed by \email, \homepage, \thanks as well.
%\collaboration{}
%\noaffiliation

%\date{\today}

\begin{abstract}
This work identifies the physical mechanisms at play in the different flow regions along an Eckart acoustic streaming jet by means of numerical simulation based on a novel modeling of the driving acoustic force including attenuation effects. The flow is forced by an axisymmetric beam of progressive sound waves attenuating over a significant part of a closed cylindrical vessel where the jet is confined. We focus on the steady, axisymmetric and laminar regime. The jet typically displays a strong acceleration close to the source before reaching a peak velocity. At further distances from the transducer, the on-axis jet velocity smoothly decays before reaching the opposite wall. For each of these flow regions along the jet, we derive scaling laws for the on-axis velocity with the magnitude of the acoustic force and the diffraction of the driving acoustic beam. These laws highlight the different flow regimes along the jet and establish a clear picture of its  spatial structure, able to inform the design of experimental or industrial setups involving Eckart streaming jets.
\end{abstract}

% insert suggested keywords - APS authors don't need to do this
%\keywords{}

%\maketitle must follow title, authors, abstract, and keywords
\maketitle

% body of paper here - Use proper section commands
% References should be done using the \cite, \ref, and \label commands

\section{Introduction} \label{section:introduction}

This study aims to characterize the jet flow produced by an axisymmetric beam of progressive, attenuated acoustic waves. We focus here on determining accurate scaling laws for the velocity along the jet in a fluid domain significantly larger than the attenuation length of the sound waves, a geometry that is representative of a number of applications of acoustically-driven flows.

Driving flows by means of acoustic waves, or acoustic streaming, relies on a nonlinear phenomenon where momentum is generated by the attenuation of sounds waves~\citep{Lighthill1978}. In this work, we focus on the Eckart-type acoustic streaming, that is a flow forced by traveling waves being attenuated in the bulk of a fluid~\cite{Eckart1948,Nyborg1953} (from now, any mention to acoustic streaming shall refer to Eckart streaming, unless specified otherwise). Because of its contactless nature, acoustic streaming has aroused considerable interest in applications in which the flow actuation by mechanical means is rendered difficult by tight geometrical constraints. This is commonly the case of microfluidics, in which the recent developments make an extensive use of sound waves to drive fluid flows and to possibly promote mixing in microcavities~\cite{Hagsater2007,Frommelt2008,Friend2011,Muller2015}, in droplets~\cite{Alghane2012,Riaud2017} and around oscillating bubbles~\cite{Cleve2019,Doinikov2022,Fauconnier2022}. Acoustic waves are also often used to trap and manipulate small objects by means of ``acoustic tweezers''~\cite{Baresch2016,Baudoin2019,Li2021}.

At a larger scale, acoustic waves constitute an appropriate way of remotely interacting with delicate fluids, for which the contamination by intrusive bodies must be avoided. This is for instance the case in metallurgy, where the access to the melts, in solidification processes, is further hindered by the important temperatures encountered and by tight thermal insulation constraints. In this context, electromagnetic~\cite{Patzold2013,Cablea2015} and mechanical~\cite{Chatelain2018} stirring of the melt have proven to be beneficial for the quality of the obtained solid phase. Eckart streaming can thus be considered as an alternative to the more intrusive actuations. It may be used to promote mixing in the liquid phase~\cite{Miralles2023,Vincent2024} or to directly act on the solid-liquid interface itself~\cite{Dridi2008a,Absar2017,ElGhani2021}. In such industrial applications, large-scale Eckart streaming flows (i.e., flows for which the characteristic length scales are significantly larger than the size of the source) are commonly obtained with high power and low frequency ultrasound waves (typically hundreds of watts and kilohertz). Due to the high acoustic wave amplitudes involved, a cloud of cavitation bubbles is produced close to the transducer and acts as the main sound attenuation mechanism driving the flow~\cite{Schenker2013,Lebon2019,Lebon2019a}. While this approach leads to strong jets, its main drawback is the loss of coherence of the sound beam over distances comparable to the size of the acoustic source. It thus restricts the controlled actuation of the fluid to regions close to the actuator. 

We are instead interested in higher frequencies and lower powers, for which the beam remains coherent over larger distances. This makes possible to act on regions of fluid located far from the acoustic transducer through the generation of elongated jets~\cite{Kamakura1996,Mitome1995,Mitome1998,Dentry2014,Moudjed2014}.  These jets can reach distances even greater than the acoustic beam attenuation length $\alpha^{-1}$, with $\alpha$ the acoustic pressure wave attenuation coefficient depending on the wave frequency and the material properties of the fluid. An important parameter for such flows is thus the quantity $N = \alpha L_s$, which compares the distance $L_s$ over which the Eckart streaming flow is observed to $\alpha^{-1}$. To ensure that most of the available energy radiated by the acoustic source is used to create flow kinetic energy, $N>1$ shall be aimed for.

Still, as most of the experimental studies focused on streaming flows generated on small distances compared to the attenuation length ($N < 1$), the current knowledge of large-scale Eckart streaming remains too scarce to efficiently implement it in an engineering context where large actuation distances are sought. From the water experiments of \citet{Kamakura1996} ($N=0.17$) and \citet{Moudjed2014} ($N=0.05$), the long streaming jet strongly accelerates over distances greater than the transducer diameter but smaller than $\alpha^{-1}$. In this region, the on-axis velocity scales with the square root of the acoustic power, as a result of a balance between inertia and the acoustic force~\citep{Moudjed2014}. At further distances from the source, \citet{Mitome1998} ($N=0.4$ and $N=0.45$) and \citet{Dentry2014} ($0.63 \leq N \leq 1400$) observed that the jet velocity reaches a maximum before decreasing further away from the source; however, the authors did not explain these jet velocity features. A global picture of the evolution of the streaming velocity with the distance from the source is thus still lacking. Aside from the cited experimental works, a coherent description of a jet obtained numerically, with justified assumptions for the modeling of the acoustic force, is scarce \cite{Baudoin2019}.

These Eckart streaming experiments call for numerical simulations to understand the scaling of streaming jets with the acoustic forcing. Direct Numerical Simulations (DNS) of the compressible Navier-Stokes equations, i.e., solving the wave propagation problem and waiting for a mean flow to build up, may appear as a natural approach to compute streaming jet flows~\citep{Benhamou2023,Daru2024}. However, as they require to solve acoustic time and length scales that are usually several orders of magnitudes smaller than those of the flow~\citep{Moudjed2014}, DNS are only tractable for streaming jet lengths that do not exceed a few wavelengths (e.g., up to 100 wavelengths in Ref.~\citep{Daru2024}). DNS are thus unsuitable for the long-range Eckart streaming we consider. Another approach relies on invoking the time scale discrepancy between the flow and acoustic problems to derive and solve a reduced set of equations describing the streaming flow~\citep{Moudjed2014,Zarembo1971,Rudenko1977}. With a method similar to RANS (Reynolds-Averaged Navier-Stokes) modeling in turbulence, \citet{Moudjed2014} showed that the streaming flow can be described by the incompressible Navier-Stokes together with an acoustic body force driving the flow. The acoustic problem then decouples from the streaming one, thus greatly alleviating the range limit encountered with DNS. Nevertheless, this approach requires the acoustic field to be accurately modeled. In particular for $N > 1$, alteration of the acoustic field structure caused by sound attenuation is expected to fundamentally change the structure of the jet. In the present work, we adapt the usual equations governing the unattenuated sound field radiated by a baffled source~\citep{Kinsler2000,Blackstock2000} to include attenuation effects while keeping a linear acoustic propagation workframe.

We present numerical simulations of Eckart streaming jets flowing in a closed cylindrical cavity significantly longer than the acoustic attenuation length, i.e., $N > 1$. This allows us to explore flow regions in which the acoustic force is either driving the flow (at distances to the transducer that are smaller than $\alpha^{-1}$) or nearly ineffective (at distances greater than $\alpha^{-1}$, where the acoustic beam is almost entirely attenuated). In the present study, we derive scaling laws for the velocity along the entire jet as a function of the forcing magnitude and the distance from the source. In particular, our analysis characterizes the velocity maximum and the downstream region of decaying velocity where the acoustic force becomes negligible.

This paper is organized as follows: first, we outline our modeling strategy in Sec.~\ref{section:problem_definition}. In particular, we define the considered problem, and provide the governing equations along with our numerical methodology. We then stress the central role of sound attenuation in Sec.~\ref{subsection:attenuation_effects}. Section~\ref{section:acoustic_streaming_jet_characterization} is devoted to the analysis of a long-range streaming flow. Finally, concluding remarks are given in Sec.~\ref{section:conclusion}.

%%%%%%%%%%%%%%%%%%%%%%%%%%%%%%%%%%%%%%%%%%%%%%%%%%%%%%%%%%%%%%%%%%%%%%%%%%%%%%%%%%%%%%%%%%
%%%%%%%%%%%%%%%%%%%%%%%%%%%%%%%%%%%%%%%%%%%%%%%%%%%%%%%%%%%%%%%%%%%%%%%%%%%%%%%%%%%%%%%%%%
%%%%%%%%%%%%%%%%%%%%%%%%%%%%%%%%%%%%%%%%%%%%%%%%%%%%%%%%%%%%%%%%%%%%%%%%%%%%%%%%%%%%%%%%%%
%%%%%%%%%%%%%%%%%%%%%%%%%%%%%%%%%%%%%%%%%%%%%%%%%%%%%%%%%%%%%%%%%%%%%%%%%%%%%%%%%%%%%%%%%%

\section{Modeling of acoustic streaming jets} \label{section:problem_definition}

\subsection{Definition of the streaming problem}

\begin{figure}[t!]
	\centering
	\includegraphics[trim={0cm, 3cm, 0.5cm, 1.5cm}, clip, scale=0.4]{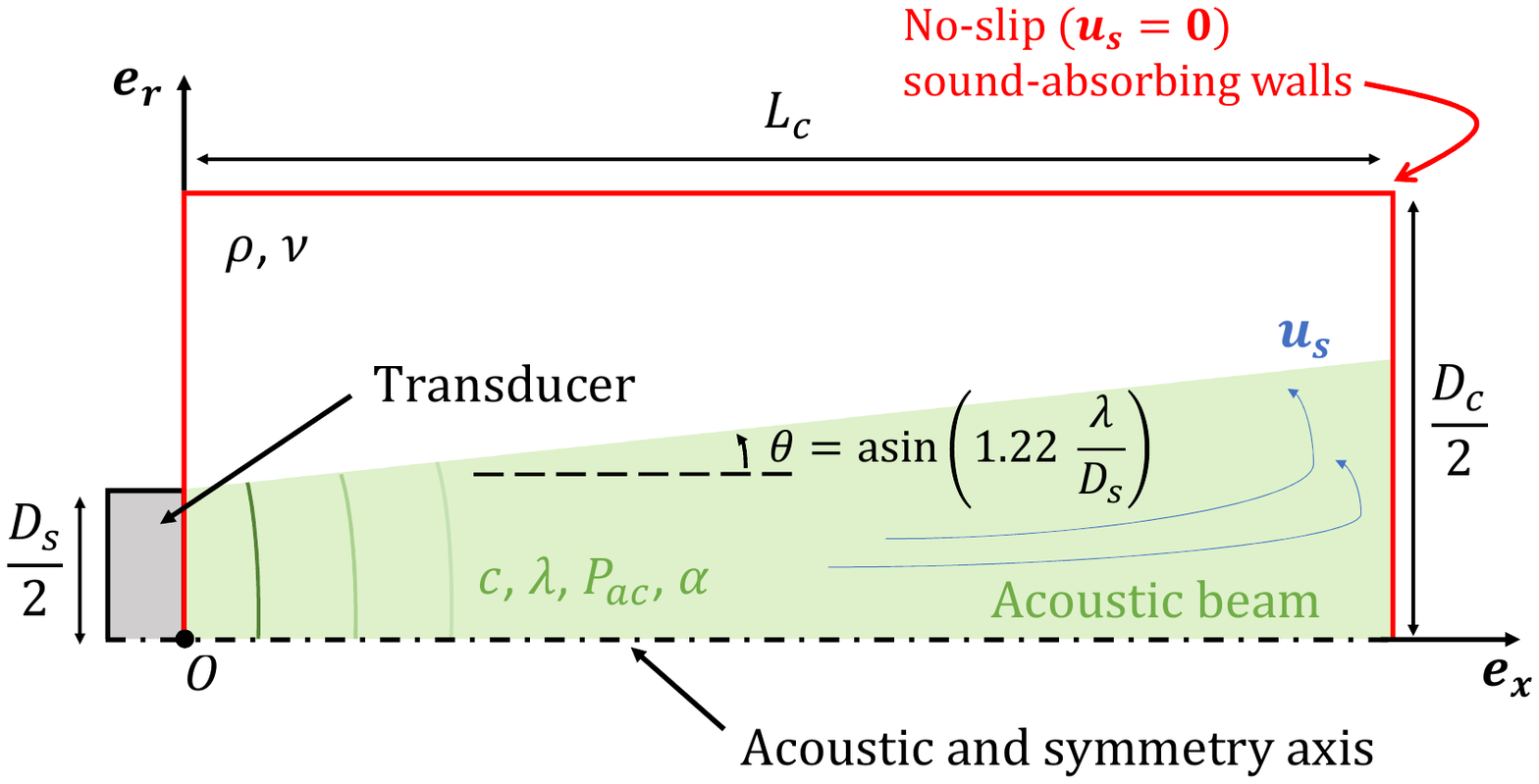}
	\caption{Sketch of the computational domain in the $\left( x_d, r_d \right)$ plane, where $x_d$ and $r_d$ refer to the dimensional axial and radial coordinates, respectively. The domain is a closed cylindrical cavity filled with a Newtonian fluid of density $\rho$ and kinematic viscosity $\nu$. It is fitted with a plane circular transducer (gray) at $x=0$ that radiates an axisymmetric beam-shaped acoustic field (green). The sound waves of wavelength $\lambda$ are emitted with an initial power $P_{ac}$ and travel at the phase speed $c$. The pressure amplitude decays at a rate $\alpha$ along the wave propagation path. The attenuation of the sound beam creates a body force that drives a flow of velocity field $\bm{u_s}$. The symmetry axis of the cylindrical container and the flow field is the $r=0$ line, and all boundaries are impermeable walls that completely absorb the sound waves.}
	\label{fig:computational_domain}
\end{figure}

We consider a steady and axisymmetric Eckart streaming flow of dimensional velocity field $\bm{u_s}$ in a closed cylindrical cavity of diameter $D_c$ and length $L_c$ (Fig.~\ref{fig:computational_domain}). The fluid is Newtonian of density $\rho$ and kinematic viscosity $\nu$.
A plane circular transducer of diameter $D_s$ is placed at $x=0$ and radiates a beam-shaped sound field centered on the cylinder axis. The sound waves, of power $P_{ac}$ and of wavelength $\lambda$, propagate through the medium at a phase speed $c$. In addition to the decay caused by diffraction, the amplitude of the acoustic pressure waves drops exponentially along the axis due to viscous and thermal effects occurring within the bulk of the fluid. This decay rate is given by the sound attenuation coefficient~\citep{Kinsler2000,Moudjed2014}:
\begin{equation} \label{eq:sound_attenuation_coefficient}
\alpha = \frac{2 \pi^2 \nu}{c \lambda^2} \left( \frac{4}{3} + \frac{\eta}{\mu} + \frac{\gamma - 1}{Pr} \right) \, ,
\end{equation}
where $\eta$ and $\mu = \rho \nu$ are the bulk and dynamic viscosities, respectively, $\gamma$ is the adiabatic index and $Pr = \mu c_p / \kappa$ is the Prandtl number with $c_p$ the specific heat capacity at constant pressure and $\kappa$ the thermal conductivity. As formerly derived by \citet{Lighthill1978}, the theoretical axial momentum flux of an Eckart streaming jet flowing in a semi-infinite domain increases with the distance $x_d$ from the transducer, and asymptotically reaches its maximum $M_{th,max} = P_{ac} / c$ at infinite $x_d$. In the case where the effective beam length matches $L_c$, only a fraction $\left( 1 - e^{- 2 \alpha L_c} \right)$ of $M_{th,max}$ is actually injected into the fluid volume. In the present work, we focus on a $L_c > \left( 1 / \alpha \right)$ case to maximize the jet momentum flux. Using $\left( \alpha, \lambda, M_{th,max}, \rho, \nu, D_s, D_c, L_c, \bm{u_s} \right)$ as dimensional design parameters, and $D_s$, $\rho D_s^3$, $D_s^2 / \nu$ as length, mass and time scales, respectively, we describe the acoustic propagation and hydrodynamics problems using six dimensionless groups~\citep{Moudjed2014}:
\begin{equation} \label{eq:dimensionless_parameters}
N = \alpha L_c \, , \quad S = 1.22 \frac{\lambda}{D_s} \, , \quad M = \frac{P_{ac}}{\rho c \nu^2} \, , \quad L = \frac{L_c}{D_s} \, , \quad D = \frac{D_c}{D_s} \, , \quad \bm{u} = \frac{ \bm{u_s} D_s}{\nu} \, .
\end{equation}
In Eq.~\eqref{eq:dimensionless_parameters}, $S$ compares the wavelength to $D_s$. As the distance from the transducer increases, the beam radius increases due to diffraction. The acoustic field thus takes the shape of a cone of half-angle $\theta = \sin^{-1} S$ (Fig.~\ref{fig:computational_domain}). In other words, the parameter $S$ expresses how diffraction acts on the shape of the acoustic force field. The Eckart streaming problem is described by the dimensionless incompressible continuity and Navier-Stokes equations:
\begin{subequations} \label{eq:incompressible_Navier_Stokes_and_continuity_dimensionless}
\begin{align}
\bm{\nabla} \bm{\cdot} \bm{u} &= 0 \, ,\label{eq:continuity_dimensionless}\\
\frac{\partial \bm{u}}{\partial t} + \left( \bm{u} \bm{\cdot} \bm{\nabla} \right) \bm{u} &= - \bm{\nabla} p + \bm{\nabla}^2 \bm{u} + Gr_{ac} \bm{ \widetilde{I} } \, . \label{eq:Navier-Stokes_dimensionless}
\end{align}
\end{subequations}
The last term in Eq.~\eqref{eq:Navier-Stokes_dimensionless} is the acoustic force driving the flow\, and is valid far from solid boundaries and for any acoustic wave shape~\citep{Riaud2017}. The spatial structure of the forcing is set by the normalized acoustic intensity $\bm{\widetilde{I}}$ dictated by the acoustic problem (see Sec.~\ref{subsection:acoustic_field}). $\bm{\widetilde{I}}$ is normalized by the maximum unattenuated plane wave intensity $I_{max}$ on the transducer axis (Eq.~\eqref{eq:acoustic_power_definition}), giving $\Vert \bm{\widetilde{I}} \Vert \leq 1$ everywhere in the fluid domain. The magnitude of the forcing, relative to the magnitude of the viscous forces, is then given by the acoustic Grashof number $Gr_{ac}$~\cite{ElGhani2021}:
\begin{equation} \label{eq:acoustic_Grashof_definition}
Gr_{ac} = \frac{32 \alpha P_{ac} D_s}{\pi \rho c \nu^2} = \frac{32}{\pi} \frac{ N M }{L} = \frac{16}{\pi} \frac{M}{L_{\alpha}}\, ,
\end{equation}
where ${L_{\alpha} = L / \left( 2 N \right) = 1 / \left( 2 \alpha D_s \right) }$ is the dimensionless attenuation length defining the exponential decay of the force along the acoustic axis. A small value of $L_{\alpha}$ corresponds to significant sound attenuation, while $L_{\alpha} \rightarrow \infty$ denotes a vanishingly small attenuation.

\begin{table}[h!]
    \caption{Values of the dimensionless parameters considered in the present study, as defined in Eqs.~\eqref{eq:dimensionless_parameters} and~\eqref{eq:acoustic_Grashof_definition}. The values of the parameters $S$ and $N / L$ correspond to the water experiments of~\citet{Moudjed2014}, for which $f = 2 \times 10^{6}$~Hz, $\alpha \approx 0.1$~m$^{-1}$, $D_s = 2.85 \times 10^{-2}$~m and $c = 1480$~m.s$^{-1}$.}
    \centering
    \begin{ruledtabular}
    \begin{tabular}{c c c c c c}
    
        $S$ & $N$ & $L$ & $D$ & $L_{\alpha} = L/(2N)$ & $Gr_{ac}$\\
        \hline
        $0.03$ & $4.5$ & $1500.7$ & $548.3$ & $166.7$ & Varied between $10^3$ and $5 \times 10^4$ \\   
        
    \end{tabular}
    \end{ruledtabular}
    \label{tab:dimensionless_parameters_values}
\end{table}

The cavity walls are modeled as impermeable boundaries, imposing $\bm{u} = \bm{0}$ at $x \in \left\{0, L \right\}$ and $r = D/2$. Besides, the flow is axisymmetric and swirl-free, so that for all $x$ and $r$:
\begin{equation*}
u_{\theta} = \frac{\partial u_x}{\partial \theta} = \frac{\partial u_r}{\partial \theta} = \frac{\partial p}{\partial \theta} = 0 \, ,
\end{equation*}
where $\theta$ is the azimuthal coordinate, and $u_x$, $u_r$ and $u_{\theta}$ refer to the axial, radial and azimuthal velocity components, respectively. Axisymmetry of both $\boldsymbol{u}$ and $p$ further imposes
\begin{equation*}
u_r = \frac{\partial u_x}{\partial r} = \frac{\partial p}{\partial r} = 0
\end{equation*}
at $r=0$.

The values of the dimensionless parameters~\eqref{eq:dimensionless_parameters} are listed in Table~\ref{tab:dimensionless_parameters_values}. We choose the same diffraction parameter $S = 0.03$ and attenuation coefficient $N/L = 3 \times 10^{-3}$ as in the water experiments of \citet{Moudjed2014}. Similar parameters values are found in the experiments of Refs.~\citep{Kamakura1996,Mitome1998,Frenkel2001}, in which $S \in \left[ 0.014, 0.036 \right]$ and $N/L \in \left[ 5.9 \times 10^{-3}, 18.8 \times 10^{-3} \right]$. The acoustic forcing structure in our simulations is thus representative of those commonly found in streaming jets experiments.

An important aspect of our work is the moderately large $N$. In contrast to Refs.~\citep{Kamakura1996,Mitome1998,Frenkel2001,Moudjed2014}, we choose $N > 1$. The beam is thus entirely attenuated before reaching the end of the fluid domain, yielding forcing-free regions at far distances form the source. Typically, at $x=L$, the remaining acoustic power is about 0.01~\% of the power radiated by the transducer. To minimize the effects of the lateral wall on the jet, we choose $D \approx L / 3$.

Finally, we take $10^3 \leq Gr_{ac} \leq 5\times 10^4$, which fits in the ranges explored by \citet{Kamakura1996,Mitome1998,Frenkel2001,Moudjed2014}. Some of these authors also investigated larger $Gr_{ac}$. We limited ourselves to $Gr_{ac} \leq 5\times 10^4$, as computing steady flows at greater $Gr_{ac}$ in such a large cavity becomes prohibitively expensive.

\subsection{Equations governing the spatial distribution of the attenuated acoustic force field}
\label{subsection:acoustic_field}

We shall now explicate the acoustic intensity field $\bm{\widetilde{I}}$, which depends only on the dimensionless group $S$ and $N/L$. As $N>1$, attenuation is expected to significantly affect $\bm{\widetilde{I}}$, so that the usual equations~\citep{Kinsler2000,Blackstock2000} for the unattenuated sound field radiated by a plane circular transducer are no longer valid and must be corrected.

Although the streaming flow is sufficiently slow to be seen as incompressible, acoustic waves are a compressible process~\citep{Chassaing2002}, and shall thus be modeled accordingly. We further assume that:
\setlist{nolistsep}
\begin{enumerate}[label=(\roman*), noitemsep]
\item Sound waves are irrotational, i.e., $\bm{\nabla} \times \bm{u_{ac}} = \bm{0}$. This is a common assumption in acoustics \citep{Kinsler2000,Blackstock2000,Coulouvrat1992}; it holds in the bulk of the fluid but breaks down in a thin layer near solid boundaries, including at the source. There, a viscous layer of dimensionless thickness $\delta = \sqrt{\nu \lambda / \left( \pi c D_s^2 \right)}$ forces the acoustic velocity fluctuation $\bm{u_{ac}}$ to match the boundary velocity. This is typically the case near the vibrating membrane of the transducer radiating the acoustic waves. For the experiment in Ref.~\citep{Moudjed2014}, on which we base our forcing field structure, $\delta \approx 1.3 \times 10^{-5}$. As $\delta$ is several orders of magnitudes smaller than the cavity and flow length scales, the supplementary attenuation and possible streaming flow within that layer can be safely neglected~\citep{Dridi2010}.
\item We assume that $L_{\alpha}$ is smaller than the acoustic shock formation distance: the sound waves are thus linear and the superposition principle applies. This modelling choice removes the need for defining a frequency~\citep{Moudjed2014} and acoustic nonlinearity parameters \citep{Coulouvrat1992}, thus making our scaling study as generic as possible.
\item The waves are time-harmonic monochromatic signals with an angular frequency $\omega$.
\item The length and time scales of the acoustic propagation problem are significantly smaller than those of the streaming flow, so that the flow does not affect sound propagation~\citep{Pierce1990}. These scale discrepancies are indeed verified in long-range laminar Eckart streaming~\citep{Kamakura1996,Mitome1998,Moudjed2014}.
\item Sound attenuation is due to the viscosity of the medium only. Asymptotic analysis~\citep{Coulouvrat1992} shows that thermally-induced attenuation is proportional to $\left( \gamma - 1 \right)$. For common liquids such as water, $\gamma \approx 1$, hence thermally-induced attenuation can be dropped~\citep{Riaud2017} (this approximation however does not hold for liquid metals). We also discard any relaxation-based attenuation, e.g., through the excitation of the vibrational and rotational modes of the fluid molecules~\citep{Kinsler2000}.
\item The sound attenuation length is large compared to the wavelength (i.e., $NS / \left(1.22 L \right) \ll 1$).
\end{enumerate}

We model the transducer as a disk vibrating at a uniform velocity $u_0 e^{j \omega t}$ (with $j^2 = -1$). That disk is modeled as a theoretically infinite number of acoustic point sources, each emitting attenuated spherical waves contributing to the acoustic pressure $\widetilde{p}_{ac}$, velocity $\bm{\widetilde{u}_{ac}}$ and $\bm{\widetilde{I}}$ at $(x, r)$ through (see appendix~\ref{appendix:acoustic_force_derivation} for the complete derivation):
\begin{subequations} \label{eq:Rayleigh_integral_all}
\begin{align}
\bm{ \widetilde{I} } &= \Re \left\{ \widetilde{p_{ac}} \, \bm{ \widetilde{u}_{ac} }^* \right\} \, , \label{eq:normalised_intensity}\\
\widetilde{p}_{ac} &= j \frac{ k^2 S }{8 \times 1.22 \pi^2} \int_{0}^{1/2} \int_{0}^{2 \pi} \frac{ e^{- j k \Vert \bm{d} \Vert } }{ \Vert \bm{d} \Vert } r' \mathrm{d}r' \mathrm{d}\theta' \, , \label{eq:Rayleigh_integral_pressure}\\
\bm{ \widetilde{u}_{ac} } &= \frac{1}{4 \pi} \int_{0}^{1/2} \int_{0}^{2\pi} \left( 1 + j k \Vert \bm{d} \Vert \right) \frac{ e^{- j k \Vert \bm{d} \Vert } }{\Vert \bm{d} \Vert^2} \frac{ \bm{d} }{ \Vert \bm{d} \Vert }  r' \mathrm{d}r' \mathrm{d}\theta' \, , \label{eq:Rayleigh_integral_velocity}\\
k &= \frac{2.44 \pi}{S} - j \frac{N}{L} \, , \label{eq:Rayleigh_integral_wavenumber}
\end{align}
\end{subequations}
where $^{*}$ refers to the complex conjugate and $\Re$ to the real part. $\bm{d} = x \, \bm{e_x} + \left[ r - r' \cos \left( \theta' \right) \right] \bm{e_r} - r' \sin \left( \theta' \right) \bm{e_{\theta}}$ is the distance between a point source located at the polar coordinates $\left( r', \theta' \right)$ from the transducer center and a point at $\left( x, r \right)$. In Eqs.~\eqref{eq:Rayleigh_integral_all}, $\widetilde{p}_{ac}$, $\bm{ \widetilde{u}_{ac} }$ and $\bm{\widetilde{I}}$ are normalized by their maximum unattenuated magnitudes, which are $2 \rho c u_0$, $2 u_0$ and $2 \rho c u_0^2$, respectively. The effects of attenuation are accounted for through the complex wavenumber $k$ (Eq.~\eqref{eq:Rayleigh_integral_wavenumber}), whose imaginary part ${N/L = 1 / \left( 2 L_{\alpha} \right) = \alpha D_s}$ is the dimensionless attenuation coefficient. Finally, since the geometry is axisymmetric, $\bm{ \widetilde{u}_{ac} } \bm{\cdot} \bm{e_{\theta}}$ vanishes when integrating Eq.~\eqref{eq:Rayleigh_integral_velocity}. Therefore, the acoustic fields are indeed axisymmetric and free of any azimuthal component.

The resulting acoustic intensity field exhibits strong gradients for $x \leq L_F$ (see Fig.~\ref{fig:transducer_mesh_convergence}~(b) and \citet{Moudjed2015}), with $L_F = 1.22 / \left( 4 S \right) \approx 10$ the Fresnel length locating the transition between the near ($x < L_F$) and far ($x > L_F$) acoustic fields. Since (i) the beam and source radii are equal to 0.5 at $x=0$ and (ii) the beam expands radially at a rate $S$ due to diffraction~\cite{Moudjed2014}, an approximate description of the beam radius variation along $x$ is:
\begin{equation} \label{eq:beam_radius}
R_{beam} = 0.5 + S x \, .
\end{equation}
The acoustic intensity~$\widetilde{I}_x(x,r=0)$ along the beam axis is decaying mainly due to this radial expansion effect but also from attenuation. The drop of $P_{ac}$ along the beam due to attenuation yields an exponential contribution to the decay of $\widetilde{I}_x(x, r=0)$. From Eq.~\eqref{eq:analytical_intensity_axis_far_field}, the length scale for this exponential decay is thus defined by $L_{\alpha} = L / \left( 2N \right) \approx 166.7$. The ratio $L_F / L_{\alpha} \approx 0.06$ is thus marginal, meaning that the near acoustic field structure is barely affected by attenuation. In our work, we choose to focus on the scaling of the jet velocity with $Gr_{ac}$ and thus keep $L_F / L_{\alpha}$ constant. For more details on the effect of $L_F / L_{\alpha}$ on the jet velocity near the source, the interested reader shall refer to \citet{Daru2024}.

\subsection{Computational methodology} \label{subsection:computational_methodology}

The problem (Eq.~\eqref{eq:incompressible_Navier_Stokes_and_continuity_dimensionless}) is solved numerically using the open source spectral-element code Semtex~\cite{Blackburn2019} on a 2D mesh~\cite{Blackburn2004}. The exponential convergence of the solution is then obtained by increasing the polynomial degree $N_p$ of the expansion basis represented at the Gauss-Lobatto-Legendre points.

\begin{figure}[h!]
	\centering
	\includegraphics[scale=0.5]{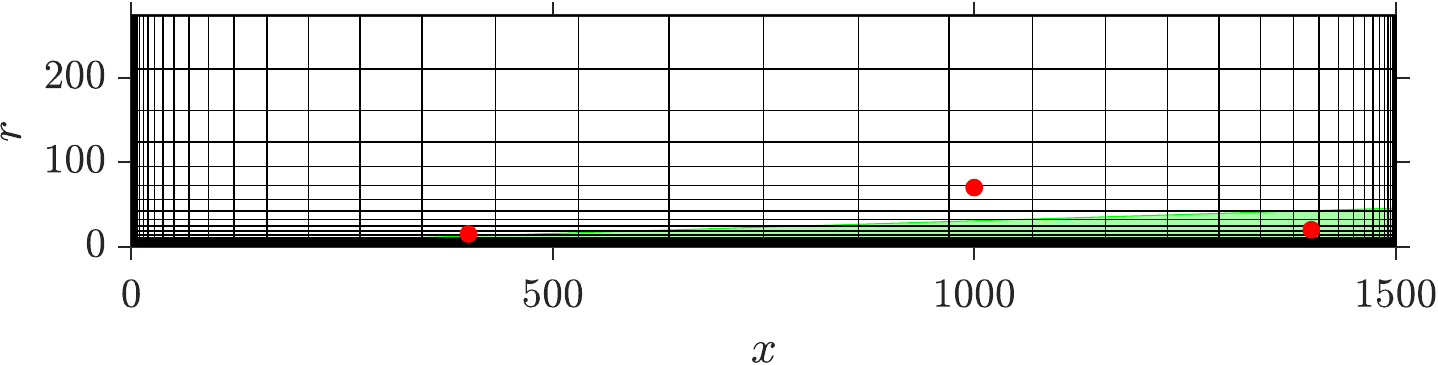}
	\caption{Discretization of the fluid domain with elements of polynomial degree $N_p = 1$. The main lobe of the unattenuated acoustic beam is represented by the green-shaded area, of approximate radius given from Eq.~\eqref{eq:beam_radius}. The red dots located at $\left( x, r \right) = \left( 400, 15 \right)$, $\left( 1000, 70 \right)$ and $\left( 1400, 20 \right)$, are the points where the time series of $\bm{u}$ and $p$ are recorded to monitor the convergence towards a steady state.}
	\label{fig:mesh_domain}
\end{figure}

The computational grid used throughout this study is shown in Fig.~\ref{fig:mesh_domain} and comprises 880 quadrilateral elements. The axial direction is discretized by 44 elements, with localized refinements at both ends of the domain~\cite{Farrashkhalvat2003} to capture the flow gradients at the jet origin and the jet impingement. The grid stretching in the $x$-direction is chosen so that the maximum ratio between the lengths of two consecutive elements does not exceed 1.35. The radial direction is discretized using 20 elements, with one element for $r \in \left[ 0, 0.5 \right]$. The element height is then continuously increased by a factor 1.3 in the ascending $r$-coordinates.

The acoustic force $Gr_{ac} \widetilde{ \bm{I} }$ is computed at the grid points prior to every flow calculations. The integrals in Eqs.~\eqref{eq:Rayleigh_integral_all} are evaluated numerically by discretizing the transducer with a uniform grid made of $N_s$ points in both azimuthal and radial directions. The total number of point sources on the transducer is thus equal to $N_s^2$.

Eqs.~\eqref{eq:Navier-Stokes_dimensionless} and~\eqref{eq:continuity_dimensionless} are initiated with $\bm{u} = \bm{0}$, and are marched forward in time using the splitting scheme of \citet{Karniadakis1991} that integrates the inertial terms explicitly and the viscous terms implicitly.
As only the steady state is of interest, the numerical scheme is used in its first-order formulation. To ensure numerical stability, the time step $\Delta t$ is adjusted for each $Gr_{ac}$ so that the maximum CFL number in the domain remains below unity. For instance, for the $Gr_{ac} = 1000$ case, $\Delta t = 10^{-3}$, yielding a maximum CFL number of 0.68. A steady state is then considered to be reached when the time series of the flow variables, recorded at different points of the domain (located by the red dots in Fig.~\ref{fig:mesh_domain}), do not show variations with time up to a precision of five significant figures.

Grid sensitivity studies were made for both the discretization of the transducer and the discretization of the fluid domain, and are given in Appendix~\ref{appendix:grid_sensitivity_analysis}. All the results presented in this document were obtained with $N_s = 300$ and $N_p = 8$. This choice of $N_s$ ensures a discrepancy of less than 1~\% between the numerical evaluation of Eqs.~\eqref{eq:Rayleigh_integral_all} for $r = 0$ and the analytical on-axis intensity profile given by Eq.~\eqref{eq:analytical_intensity_acoustic_axis_dimensionless}. Regarding the discretization of the fluid domain, $N_p = 8$ yields a marginal error of only about 0.01~\% on both the location and the amplitude of the maximum on-axis jet velocity.

\section{The role of the acoustic force attenuation}
\label{subsection:attenuation_effects}

In the literature, the acoustic attenuation has frequently been discarded in the model of the acoustic force to derive analytical 1D-1C (one-dimensional, one-component) Eckart streaming velocity profiles within a weak attenuation assumption~\citep{Eckart1948,Nyborg1953,Rudenko1977,Rudenko1998,Nyborg1998,Dridi2008a,Dridi2010,BenHadid2012}. We want to emphasize the importance of the contribution of the attenuation in the spatial structure of the acoustic force, even for cases where the jet observation length is small compared to $L_{\alpha}$. For this purpose we will focus on numerical calculations of the on-axis velocity profile, the radius of the jet and the momentum flux of the jet, by computing Eqs.~\eqref{eq:Rayleigh_integral_all} considering two cases: $N/L = 0$ in Eq.~\eqref{eq:Rayleigh_integral_wavenumber} (without attenuation, as it is often considered in the literature) and $N/L \neq 0$ (with attenuation, as it will be considered from Sec.~\ref{section:acoustic_streaming_jet_characterization} onwards). We shall then compare the numerical results to the theoretical predictions of Lighthill~\citep{Lighthill1978} regarding the jet momentum flux.

\begin{figure}[h!]
	\centering
	\includegraphics[scale=0.57]{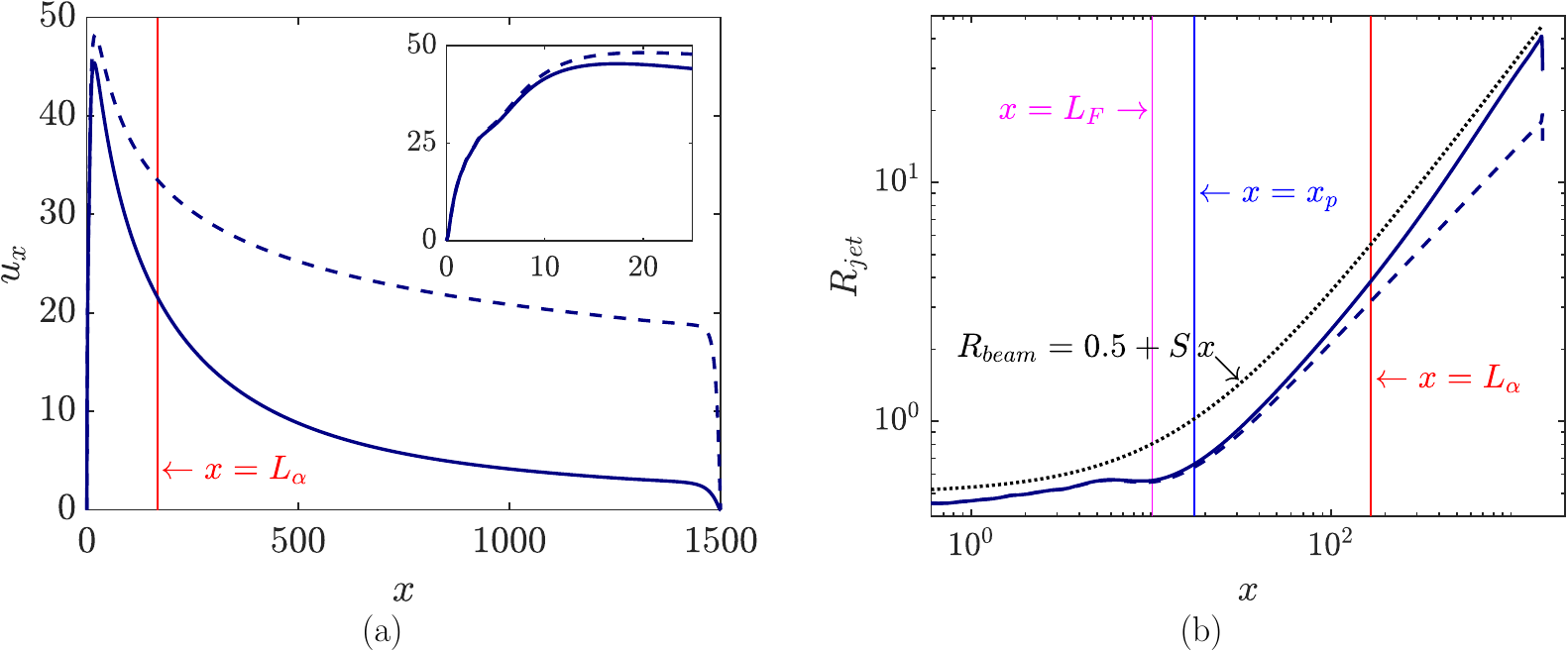}
	\caption{Effects of acoustic force attenuation in the expression of $\widetilde{ \bm{I} }$ (Eqs.~\eqref{eq:Rayleigh_integral_all}) on steady flows computed for $Gr_{ac} = 10^3$. The cases of attenuated ($N/L=0.003$) and unattenuated ($N/L=0$) $\bm{\widetilde{I}}$ are plotted with solid and dashed lines, respectively. (a) Longitudinal profiles of the on-axis velocity $u_x$, with a detailed view close to the transducer shown in the inset. (b) Longitudinal profiles of the jet radius $R_{jet}$ defined as $u_x \left( x, R_{jet} \right) = 0.5 \, u_x \left(x, 0 \right)$. The approximate beam radius (Eq.~\eqref{eq:beam_radius}) is shown as a black dotted curve. The purple and red vertical lines locate the Fresnel distance $L_F$ and the force attenuation distance $L_{\alpha}$, respectively. In (b), the vertical blue line locates the on-axis velocity peak position $x_p$ in the attenuated $\bm{\widetilde{I}}$ case.}
	\label{fig:attenuation_effects_velocity_radii_profiles}
\end{figure}

We first compare on Fig.~\ref{fig:attenuation_effects_velocity_radii_profiles}~(a), the on-axis jet velocity at $Gr_{ac} = 10^3$ for the model including attenuation (solid line) and discarding attenuation (dashed line). For $x < 8$ (i.e., $x / L_{\alpha} < 5 \times 10^{-2}$), no significant difference between the two velocity profiles is observed. However, the discrepancy builds up further downstream. In the absence of attenuation in $\bm{\widetilde{I}}$, $u_x$ reaches a maximum at a greater distance from the transducer, and with a larger magnitude. The velocity drop downstream the peak is also greatly affected, as $u_x$ decays at a significantly slower rate with $x$. In the vicinity of the downstream wall, the observed velocity magnitudes typically differ by a factor 10 between the attenuated and non-attenuated cases.

The large discrepancy observed between the on-axis velocity profiles calculated with and without attenuation of $\bm{ \widetilde{I} }$ is accompanied by a different radial spreading rate of the jet (Fig.~\ref{fig:attenuation_effects_velocity_radii_profiles}~(b)). In the near-acoustic field region ($x \leq L_F$), the jet radius $R_{jet}$ (defined as the radial coordinate where $u_x$ is 50~\% of the on-axis velocity) remains nearly constant for both the attenuated and unattenuated cases. Beyond $L_F$, $R_{jet}$ increases similarly to $R_{beam}$ and up to $x \approx 2 \times 10^2$ for an unattenuated $\widetilde{ \bm{I} }$, whereas $R_{jet}$ experiences a larger increase in the attenuated case. This difference is related to the jet velocity: when high (as for the unattenuated $\bm{ \widetilde{I} }$), viscous diffusion weakly affects the radial spread of the jet. Thus, the shape of the jet tends to follow more closely the shape of the beam, which is, in turn, dictated by acoustic diffraction. By contrast, low velocities (as for an attenuated $\widetilde{ \bm{I} }$), increase the role of viscous diffusion. This leads to a greater radial enlargement of the jet as $x$ increases, as observed in Fig.~\ref{fig:attenuation_effects_velocity_radii_profiles}~(b).

\begin{figure}[h!]
	\centering
	\includegraphics[trim={0cm, 0, 0cm, 0}, clip, width=0.95\textwidth]{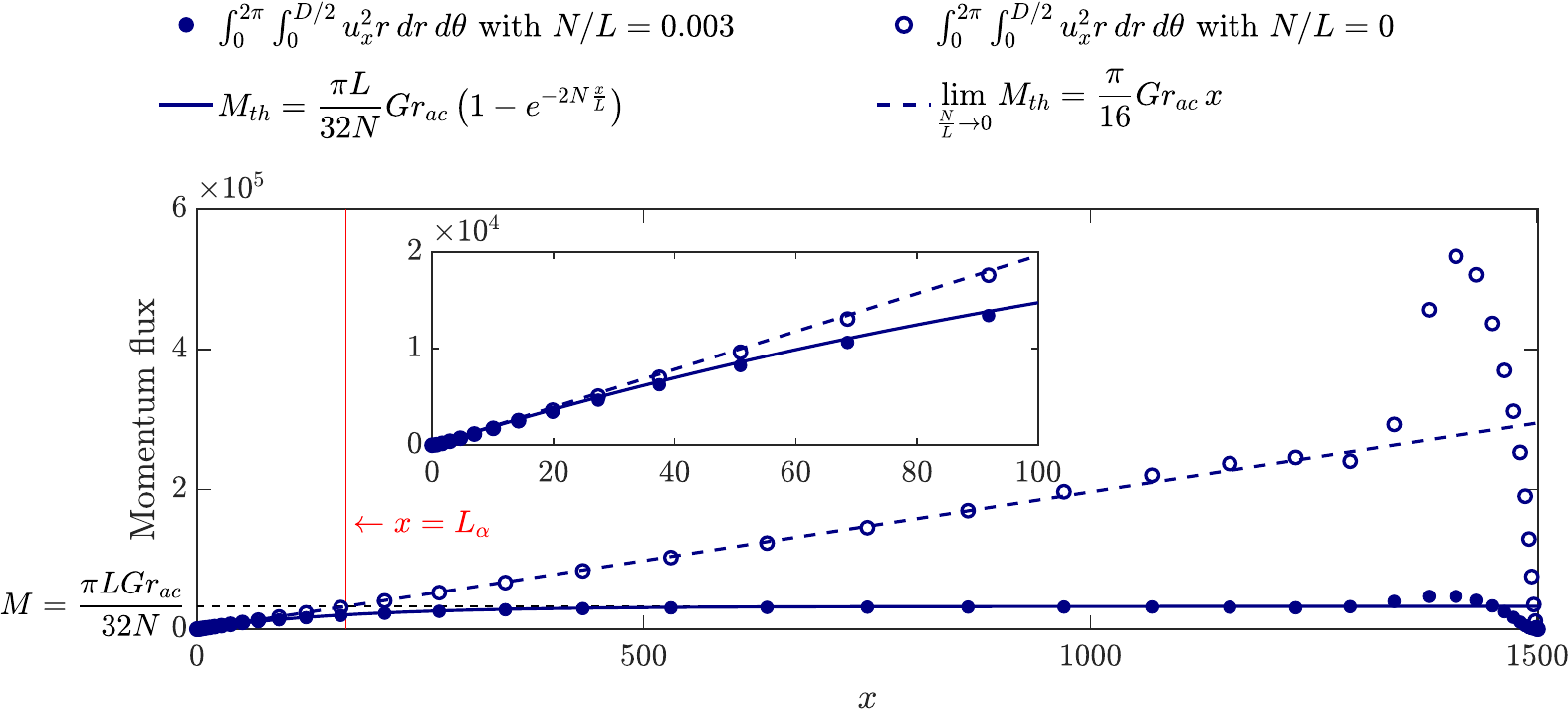}
	\caption{Evolution of the momentum flux accumulated by the jet with the distance $x$ from the transducer for $Gr_{ac} = 10^3$. The numerical results obtained when attenuation is accounted for ($N / L = 0.003$, Table~\ref{tab:dimensionless_parameters_values}) or not ($N / L = 0$) in $\widetilde{ \bm{I} }$ (Eqs.~\eqref{eq:Rayleigh_integral_all}) are represented by the filled and empty symbols, respectively. The numerical results are overlayed to the profiles of the injected momentum flux $M_{th}$, given by Eq.~\eqref{eq:axial_momentum_theoretical} and evaluated for $N / L = 0.003$ (solid blue line) and for $N / L \rightarrow 0$ (dashed blue line). The red vertical line locates the force attenuation length $L_{\alpha} = L / \left( 2 N \right)$. The horizontal black dashed line indicates the maximum momentum flux $M$ (from~\citet{Lighthill1978}, Eq.~\eqref{eq:dimensionless_parameters}). A detailed view of the momentum profiles, closer to the transducer, is shown in the inset.}
	\label{fig:attenuation_effects_momentum_profiles}
\end{figure}

From these numerical results, we can now study the evolution along $x$ of the axial jet momentum flux defined as $\int_{0}^{2 \pi} \int_{0}^{D/2} u_x^2 r dr d\theta$. As illustrated in Fig.~\ref{fig:attenuation_effects_momentum_profiles} for $Gr_{ac} = 10^3$, when $N / L \neq 0$, the axial jet momentum flux (filled symbols) saturates for $x \geq 500$ up to $x = 1300$. On the contrary, setting $N / L = 0$ leads to a linear increase with $x$ of the axial momentum flux (empty symbols). This is in agreement with the theoretical prediction by \citet{Lighthill1978} for an axisymmetric Eckart streaming jet driven by an attenuated sound beam in a semi-infinite medium: the evolution with $x$ of the axial momentum flux, then defined as $\int_{0}^{2 \pi} \int_{0}^{\infty} u_x^2 r dr d\theta$, is equal to the integral of the acoustic force along the beam if the pressure gradient is neglected. This momentum injected by the acoustic forcing in the fluid will be denoted $M_{th}$, and obeys to the following expression: 
\begin{equation} \label{eq:axial_momentum_theoretical}
M_{th} =  M \left( 1 - e^{-2 N \frac{x}{L} } \right) =  M \left( 1 - e^{-\frac{x}{L_{\alpha}} } \right)\, .
\end{equation}
For $x / L_{\alpha} \gg 1$, the injected momentum flux saturates towards the value $M$. The corresponding maximum dimensional momentum flux to be acquired by the jet is then $P_{ac} / c$. On the other hand, in the limit of vanishing attenuation, for $x / L_{\alpha} \ll 1$, Eq.~\eqref{eq:axial_momentum_theoretical} reduces to $\displaystyle{\lim\limits_{\frac{N}{L} \to 0} M_{th} \simeq \frac{M x}{L_{\alpha}} = \pi Gr_{ac} x / 16}$. In other words, beam attenuation directly dictates a length scale $L / \left( 2 N \right) = L_{\alpha}$ over which the jet momentum flux is created, and ignoring attenuation in the acoustic force leads to an unimpeded build-up of the momentum flux along the jet. These theoretical predictions are well recovered in the flow simulations (Fig.~\ref{fig:attenuation_effects_momentum_profiles}): for $x < 1100$, the theoretical and numerical profiles differ by less than 3~\%. For $x > 1300$, the larger discrepancies for both the attenuated and unattenuated cases are due to the recirculation imposed by the downstream wall, which is absent in the free-jet setting of \citet{Lighthill1978}. Finally, for very small values of $x$ (inset of Fig.~\ref{fig:attenuation_effects_momentum_profiles}), the discrepancy between the two numerical momentum flux profiles is weak. Nevertheless, the relative error $\epsilon$ on the momentum flux increases rapidly with $x$: for instance, $\varepsilon \approx 0.05$ for $x = 0.1 L_{\alpha}$ and $\varepsilon \approx 0.6$ for $x = L_{\alpha}$. Hence, discarding the effects of attenuation in
$\widetilde{ \bm{I} }$ greatly affects the flow, even at $x\ll L_{\alpha}$.

To conclude, the maximum portion of acoustic power used to create momentum flux is limited to $P_{ac} / c$ which does not depend on $\alpha$. Attenuation of the beam thus only sets the length scale over which the momentum flux is created, not its maximum value; and only a fraction of $P_{ac}$ generates momentum. Neglecting the attenuation of $\bm{\widetilde{I}}$, even when weak, results in large errors on the velocity of the jet and on its radial enlargement, even at short range ($x \ll L_{\alpha}$). It is then of primary importance to include the attenuation while modeling the acoustic force even for small ${x /L_{\alpha}}$. From now, all the results are obtained for an attenuated expression of $\bm{\widetilde{I}}$.

%%%%%%%%%%%%%%%%%%%%%%%%%%%%%%%%%%%%%%%%%%%%%%%%%%%%%%%%%%%%%%%%%%%%%%%%%%%%%%%%%%%%%%%%%%
%%%%%%%%%%%%%%%%%%%%%%%%%%%%%%%%%%%%%%%%%%%%%%%%%%%%%%%%%%%%%%%%%%%%%%%%%%%%%%%%%%%%%%%%%%
%%%%%%%%%%%%%%%%%%%%%%%%%%%%%%%%%%%%%%%%%%%%%%%%%%%%%%%%%%%%%%%%%%%%%%%%%%%%%%%%%%%%%%%%%%
%%%%%%%%%%%%%%%%%%%%%%%%%%%%%%%%%%%%%%%%%%%%%%%%%%%%%%%%%%%%%%%%%%%%%%%%%%%%%%%%%%%%%%%%%%

\section{Characterization of long-range acoustic streaming jets}
\label{section:acoustic_streaming_jet_characterization}

\begin{figure}[h!]
	\centering
	\includegraphics[trim={0, 4cm, 0, 4cm}, clip, scale=0.6]{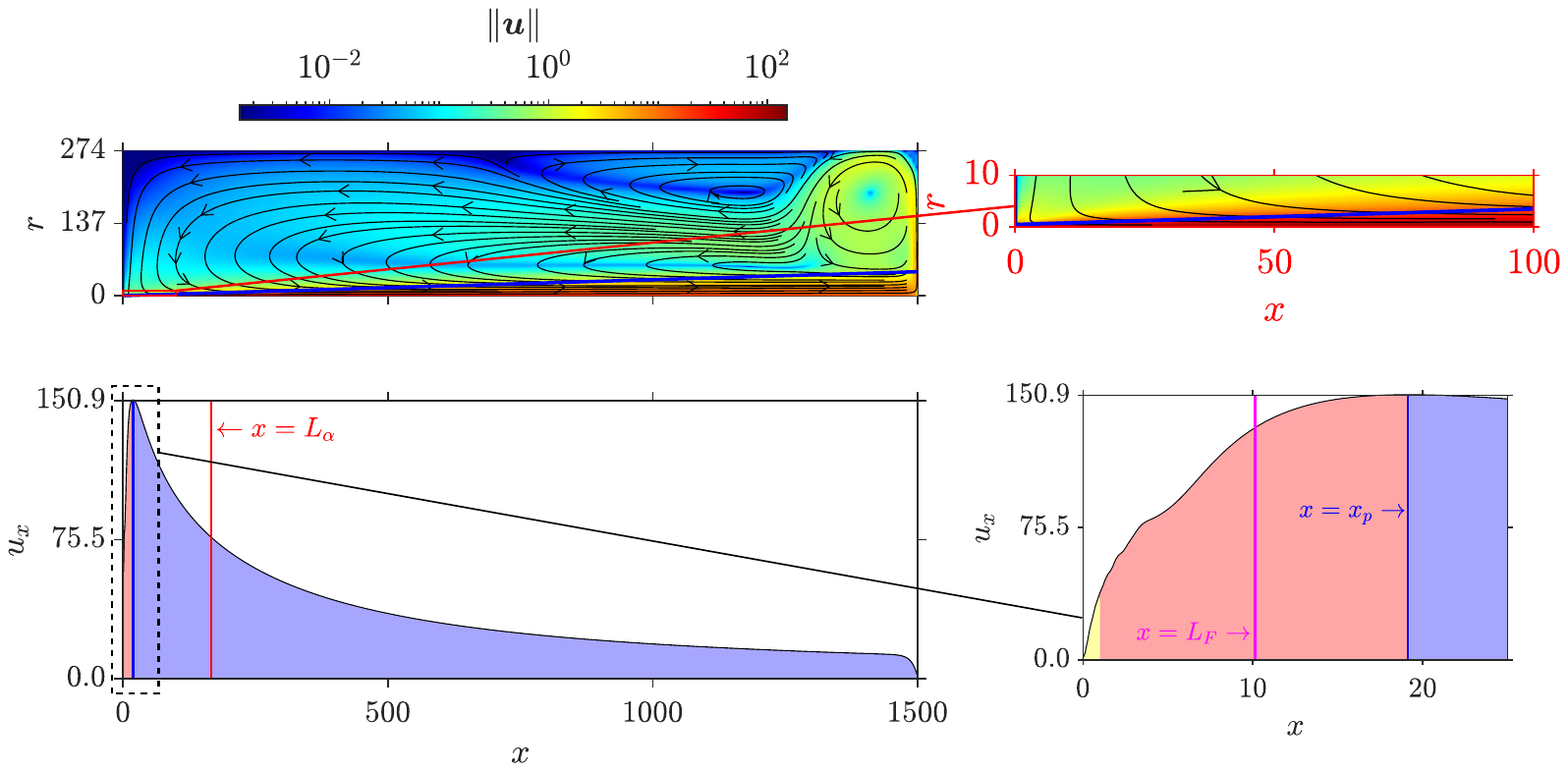}
	\caption{Typical steady velocity field obtained for $Gr_{ac} = 5 \times 10^3$. Top: Map of velocity magnitude together with streamlines. The approximate beam radius, defined by Eq.~\eqref{eq:beam_radius}, is displayed in blue, and the inset on the right focuses on the flow region close to the transducer. Bottom: longitudinal profile of $u_x$ on the jet axis on which the Fresnel distance $L_F = 1.22 / \left( 4 \, S \right)$, the on-axis velocity peak position $x_p$ and the acoustic force attenuation distance $L_{\alpha} = L / \left( 2 \, N \right)$ are reported. The right plot focuses on the region framed in black in the left plot.}
	\label{fig:velocityField_GrAc5000}
\end{figure}

We shall now thoroughly characterize the streaming jet. A typical illustration of the studied velocity fields is shown in Fig.~\ref{fig:velocityField_GrAc5000}.
The flow is steady, axisymmetric and laminar. The evolution of the jet velocity along its axis is characterized by a strong acceleration near the transducer (yellow and red regions in the bottom plot of Fig.~\ref{fig:velocityField_GrAc5000}). The irregularities witnessed in the on-axis velocity profile for $1 \leq x \leq 10$ are caused by the large longitudinal gradients of the force in the acoustic near field (Fig.~\ref{fig:transducer_mesh_convergence}~(b)) and were discussed in previous works~\cite{Kamakura1996,Moudjed2015}. On the axis, the jet velocity then reaches a peak value, whose position $x_p$ does not coincide with either of the acoustic quantities $L_F$ or $L_{\alpha}$. At $x_p$, the Reynolds number is maximum and is equal to 150.9 for $Gr_{ac} = 5 \times 10^3$. Downstream $x_p$, the on-axis jet velocity decays smoothly as the distance from the transducer increases (blue region in the bottom plot of Fig.~\ref{fig:velocityField_GrAc5000}). Despite flowing over large distances compared to $L_{\alpha}$ (i.e., up to distances over which the sound beam has been nearly entirely attenuated), the jet impinges the downstream wall with a velocity of nearly 10~\% of its peak value. The jet impingement occurring at $x = L$ then gives rise to a large vortex of weak velocity amplitude. In the following, we shall only consider the jet itself and each of the flow features identified on the axial jet velocity profile, from the fluid region close to the transducer up to the downstream wall. We shall rely on an order-of-magnitude approach~\citep{Bejan2013} (i) to determine the approximate magnitude of the on-axis velocity $u_x$ in the jet regions identified in the bottom plots of Fig.~\ref{fig:velocityField_GrAc5000}, (ii) to explain the evolution of $u_x$ in these regions based on a force balance, and (iii) to determine how $u_x$ scales with the parameters of the problem.

\subsection{Inertia-dominated region of accelerating fluid} \label{subsection:inertial_regime}

We shall now investigate the changes of jet velocity when $Gr_{ac}$ is varied from $10^3$ to $5\times 10^4$ while keeping all other parameters constant. We shall first focus on the $0.5 \leq x \leq x_p$ region over which the jet accelerates. We will not consider the region  $x \leq 0.5$ (shown in yellow in the inset of the bottom plot in Fig.~\ref{fig:velocityField_GrAc5000}): it was already studied in Ref.~\cite{Moudjed2015} and is extremely small compared the length of the fluid domain.

\begin{figure}[h!]
	\centering
	\includegraphics[width=0.98\textwidth]{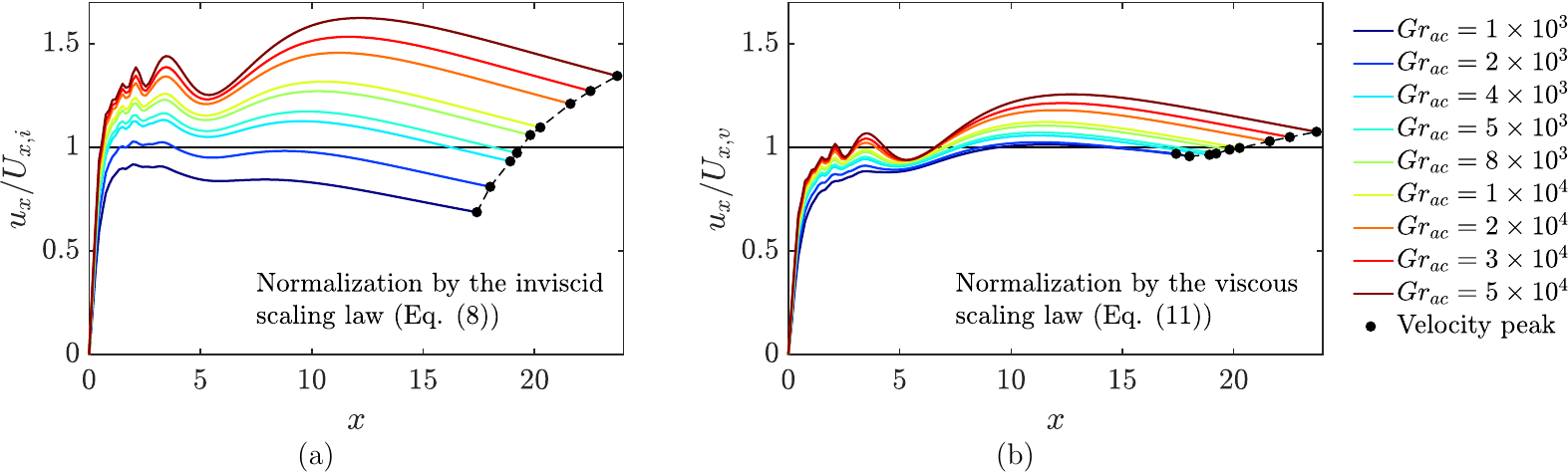}
	\caption{Longitudinal profiles of the on-axis jet velocity normalized by different scaling laws, from the source ($x=0$) to the velocity peak position $x_p$ (black points). The profiles are shown for different acoustic Grashof numbers $Gr_{ac}$. (a) On-axis velocity normalized by the inviscid scaling law $U_{x,i}$ (Eq.~\eqref{eq:OM_inertial_Moudjed},~\cite{Moudjed2014}). (b) On-axis velocity normalized by the improved scaling law accounting for viscosity effects (Eq.~\eqref{eq:OM_inertial_corrected}). As the $u_x$ is normalized in both cases by the derived scaling law, an exact description of the profile would appear as a straight horizontal line at a unit ordinate.}
	\label{fig:inertialRegimeScaling}
\end{figure}

For $0.5 \leq x \leq x_p$, former studies have shown that the jet accelerates due to a balance between the acoustic and inertia forces~\cite{Moudjed2014,Orosco2022}. In these previous works, the authors considered that since the length of the acceleration region is small compared to $L_{\alpha}$, it was legitimate to neglect both the attenuation and the beam diffraction~\citep{Daru2024}. The order of magnitude $U_{x,i}$ of the inviscid on-axis velocity was then given by~\cite{Moudjed2014}:
\begin{equation} \label{eq:OM_inertial_Moudjed}
U_{x,i} \sim \frac{1}{2} \sqrt{ Gr_{ac} \, x} \, .
\end{equation}
Equation~\eqref{eq:OM_inertial_Moudjed} fairly describes the increase of $u_x$ on $0.5 \leq x \leq x_p$ as can be seen on Fig.~\ref{fig:inertialRegimeScaling}~(a), in which an exact description of $u_x(x,r=0)$ would give a straight horizontal line at ${u_x / U_{x,i} = 1 }$. However, it either overestimates $u_x$ by nearly 30~\% for $Gr_{ac} = 10^3$, or instead underestimates it by 55~\% for $Gr_{ac} = 5 \times 10^4$. This discrepancy on the dependency on $Gr_{ac}$ between Eq.~\eqref{eq:OM_inertial_Moudjed} and the numerical profiles may be attributed to viscosity effects, neglected in Ref.~\cite{Moudjed2014}.
An order of magnitude accounting for viscous forces can be obtained from the first component of the Navier-Stokes equation, Eq.~\eqref{eq:Navier-Stokes_dimensionless}. On $0.5 \leq x \leq x_p$, the jet is almost purely axial and the order of magnitude of the inertia term in Eq.~\eqref{eq:Navier-Stokes_dimensionless} is given by $u_x \partial u_x / \partial x$ \citep{Moudjed2014}. Furthermore, the pressure forces are negligible due to the large cavity diameter $D$. From an order of magnitude standpoint, the Navier-Stokes equations~\eqref{eq:Navier-Stokes_dimensionless} thus give:
\begin{equation} \label{eq:inertial_regime_balance_viscosity}
\frac{1}{2} \frac{ \partial \left( U_{x,v}^2 \right) }{\partial x} \simeq - \frac{U_{x,v}}{R_{jet}^2} + Gr_{ac} \widetilde{I}_x \, ,
\end{equation}
where $U_{x,v}$ is an order of magnitude of $u_x\left(x, r=0 \right)$ accounting for viscous forces, and $\widetilde{I}_x = \widetilde{ \bm{I} } \cdot \bm{e_x}$. From the acoustic power definition (Eq.~\eqref{eq:acoustic_power_definition}) and its attenuation along a beam of varying cross-section:
\begin{equation} \label{eq:OM_intensity}
\widetilde{I}_x \sim \frac{ e^{- \frac{x}{L_{\alpha}} } }{16 R_{beam}^2} \, .
\end{equation}
Since both diffraction and attenuation are weak over the jet region of interest (i.e., $S \ll 1$ and $x \ll L_{\alpha}$ respectively), the spatial variations of $\widetilde{I}_x$ caused by attenuation and diffraction can be neglected in Eq.~\eqref{eq:OM_intensity}. Besides, Fig.~\ref{fig:attenuation_effects_velocity_radii_profiles}~(b) shows that $R_{jet} \sim R_{beam} \sim 1 / 2$ near the transducer. Further taking $\frac{\partial \left( U_{x,v}^2 \right)}{\partial x} \sim U_{x,v}^2 / x$, Eq.~\eqref{eq:inertial_regime_balance_viscosity} then reduces to a quadratic polynomial of positive root
\begin{equation} \label{eq:OM_inertial_corrected}
U_{x,v} \simeq - 4 x + \frac{ \sqrt{2} }{2} \sqrt{ 32 x^2 + Gr_{ac} x } \, .
\end{equation}
Equation~\eqref{eq:OM_inertial_corrected} improves on Eq.~\eqref{eq:OM_inertial_Moudjed} as it incorporates the effect of viscous forces while still neglecting acoustic attenuation and diffraction. That new scaling for $U_{x,v}$ is compared to the numerical data in Fig.~\ref{fig:inertialRegimeScaling}~(b) in which, as in Fig.~\ref{fig:inertialRegimeScaling}~(a), an exact description of $u_x(x,r=0)$ would give a straight horizontal line at ${u_x / U_{x,v} = 1} $. A glance at this plot is enough to conclude that Eq.~\eqref{eq:OM_inertial_corrected} significantly improves the prediction of the on-axis velocity profile $u_x(x,r=0)$ upstream the velocity peak: the discrepancy between $U_{x,v}$ and $u_x(x,r=0)$ is reduced to at most 25~\%, confirming that the viscous forces play a non-negligible role in the $0.5 \leq x \leq x_p$ jet region. We consider this a fair improvement in the frame of the order-of-magnitude approach adopted here.

\subsection{Scaling of the on-axis velocity peak position and amplitude} \label{subsection:velocity_peak}

The region of jet acceleration is followed by a velocity peak $u_{x,p}$ at $x_p$. Both $x_p$ and $u_{x,p}$ display a weak dependency on $Gr_{ac}$ that we shall now investigate. At $x=x_p$ and $r=0$, both $\partial u_x / \partial x$ and $\partial u_x / \partial r$ are zero, causing the left-hand side of the steady form of the Navier-Stokes equation~\eqref{eq:Navier-Stokes_dimensionless} to vanish. For any $Gr_{ac}$, $u_{x,p}$ is thus controlled by a local balance between the acoustic and viscous forces:
\begin{equation} \label{eq:momentum_balance_peak}
2 \frac{U_{x,p}}{R_{jet,p}^2} \sim Gr_{ac} \widetilde{I}_{x}(x,r=0) \, ,
\end{equation}
where $U_{x,p}$ and $R_{jet,p}$ are orders of magnitude of the velocity amplitude and jet radius at $x_p$, respectively. Based on the definition of the axial momentum flux and Eqs.~\eqref{eq:axial_momentum_theoretical}, $R_{jet,p}$ can be related to $U_{x,p}$ through:
\begin{equation} \label{eq:OM_axial_momentum}
U_{x,p}^2 R_{jet,p}^2 \sim \frac{L_{\alpha}}{32} Gr_{ac} \left( 1 - e^{ - \frac{x_p}{L_{\alpha}} } \right) \, .
\end{equation}
We thus assume that $D$ is sufficiently large so that the sidewall does not affect $x_p$. In the present work, the on-axis velocity peak occurs in the acoustic far field where the acoustic intensity asymptotically evolves as $\widetilde{I}_x(x,r=0) \sim \left( \frac{1.22 \pi}{8 \, S} \right)^2 \frac{ e^{- \frac{x}{L_{\alpha}} } }{x^2} $ (see Eqs.~\eqref{eq:analytical_intensity_axis_far_field}). Assuming $x_p \ll L_{\alpha}$, both that far-field expression of $\widetilde{I_x}$ and Eq.~\eqref{eq:OM_axial_momentum} can be simplified using Taylor's expansions so that the viscous $vs$ acoustic force balance~\eqref{eq:momentum_balance_peak} becomes:
\begin{equation} \label{eq:velocity_peak_amplitude}
U_{x,p} \sim \left( \frac{ 1.22 \pi \, Gr_{ac} }{64 \, S} \right)^{\frac{2}{3}} \, x_p^{-\frac{1}{3}} \, .
\end{equation}
\begin{figure}[t!]
	\centering
	\includegraphics[width=0.98\textwidth]{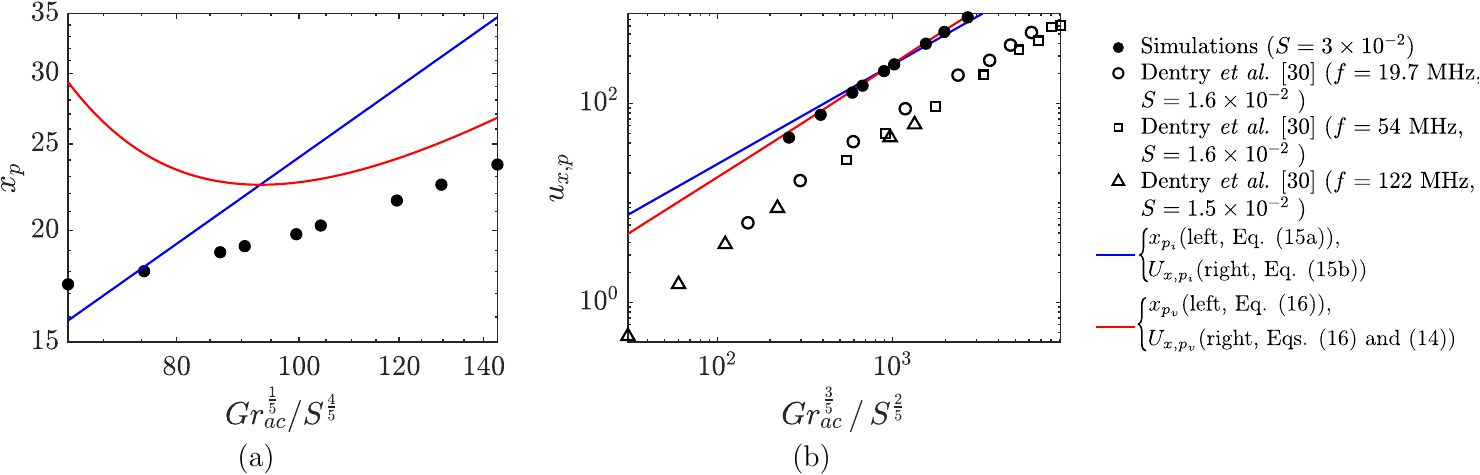}
	\caption{Evolution of the on-axis velocity peak position $x_p$ and amplitude $u_{x,p}$ with $Gr_{ac}$. (a) Comparison between the peak positions observed in the numerical simulations (black filled symbols), and the predictions when neglecting  viscous forces ($x_{p,i}$, blue line, Eq.~\eqref{eq:velocity_peak_position_inviscid}) and considering them ($x_{p,v}$, red line, Eq.~\eqref{eq:velocity_peak_position_viscous}) in the region upstream $x_p$. As the velocity peak lies in the far-acoustic field, both $x_{p,i}$ and $x_{p,v}$ are determined using the far-field approximation~\eqref{eq:analytical_intensity_axis_far_field} for the on-axis intensity $\widetilde{I}_x$. The quantity represented on the $x$-axis is chosen to assess the validity of Eq.~\eqref{eq:velocity_peak_position_inviscid}. (b) Comparison between the peak amplitudes observed in the numerical simulations (black filled symbols), the experiments from \citet{Dentry2014} (empty symbols) and the theoretical predictions (solid lines) when viscous forces upstream $x_p$ are neglected ($U_{x,p_i}$, blue, Eq.~\eqref{eq:velocity_peak_amplitude_inviscid}) and taken into account ($U_{x,p_v}$, red, Eqs.~\eqref{eq:velocity_peak_position_viscous} and~\eqref{eq:velocity_peak_amplitude}). The quantity plotted on the $x$-axis is chosen to assess the validity of Eq.~\eqref{eq:velocity_peak_amplitude_inviscid}. In (b), the data of \citet{Dentry2014} correspond to a jet of rectangular cross section. The beam driving the jet is radiated by a rectangular Surface Acoustic Waves (SAW) device of height $H_s$ that varies with frequency. For the data points of \citet{Dentry2014}, $Gr_{ac}$ and $S$ are thus based on $H_s$ (instead of $D_s$ in our numerical simulations).}
	\label{fig:velocity_peak_position_and_amplitude}
\end{figure}
Finally, substituting the purely inertial velocity estimate $U_{x,i}$ (Eq.~\eqref{eq:OM_inertial_Moudjed}) for $U_{x,p}$ in Eq.~\eqref{eq:velocity_peak_amplitude} yields:
\begin{subequations} \label{eq:velocity_peak_position_and_amplitude_inviscid}
\begin{gather}
x_{p_i} \sim \left[ \left( \frac{1.22 \pi}{8 \, S} \right)^4 \frac{ Gr_{ac} }{64} \right]^{ \frac{1}{5} } , \label{eq:velocity_peak_position_inviscid} \\
U_{x,p_i} \sim \frac{1}{2} \left[ \frac{1}{64} \left( \frac{1.22 \pi}{8 \, S} \right)^4 \right]^{ \frac{1}{10} } Gr_{ac}^{ \frac{3}{5} }  \label{eq:velocity_peak_amplitude_inviscid} \, .
\end{gather}
\end{subequations}
Equation~\eqref{eq:velocity_peak_position_inviscid} is plotted as a blue line in Fig.~\ref{fig:velocity_peak_position_and_amplitude}~(a). Neglecting the viscosity effects in the fluid region upstream the velocity peak is seen to lead to overestimate the increase of $x_p$ with $Gr_{ac}$ observed in the simulations (represented as filled symbols). This leads to an error on $x_p$ ranging from 1~\% to 46~\% over the $Gr_{ac}$ range. 

As seen in section~\ref{subsection:inertial_regime}, for higher values of $Gr_{ac}$, Eq.~\eqref{eq:OM_inertial_corrected} is expected to give a better estimate of $U_{x,p}$ than Eq.~\eqref{eq:OM_inertial_Moudjed}, then an estimate of the peak position $x_{p_v}$ is obtained numerically from:
\begin{equation} \label{eq:velocity_peak_position_viscous}
4 x_{p_v}^{ \frac{4}{3} } - \frac{ \sqrt{2} }{2} \left( 32 x_{p_v}^{ \frac{8}{3} } + Gr_{ac} x_{p_v}^{ \frac{5}{3} } \right)^{ \frac{1}{2} } + \left( \frac{1.22 \pi \, Gr_{ac}}{64 \, S} \right)^{ \frac{2}{3} } = 0 \, .
\end{equation}
The corresponding velocity magnitude $U_{x,p_v}$ at $x_{p_v}$ is then evaluated using either Eq.~\eqref{eq:velocity_peak_amplitude} or~\eqref{eq:OM_inertial_corrected} together with the obtained value of $x_{p_v}$.

The prediction $x_{p_v}$ (red curve in Fig.~\ref{fig:velocity_peak_position_and_amplitude}~(a)) yields a better estimate of the peak position for $Gr_{ac}^\frac{1}{5} / S^{\frac{4}{5}} \geq 105$ (i.e., $Gr_{ac} \geq 10^4$). At lower $Gr_{ac}$, the larger discrepancy is caused by the dependency of $R_{jet}$ on $Gr_{ac}$. We recall that $x_{p_v}$ relies on the velocity scaling law (\ref{eq:OM_inertial_corrected}) obtained by assuming that $R_{jet} \approx 0.5$ i.e., the jet width is set by the beam width for all $Gr_{ac}$. While this approximation holds for $Gr_{ac} \geq 10^4$, we however observed that the increased viscous diffusion at lower $Gr_{ac}$ widens the jet. For instance, $R_{jet}$ measured at $x = L_F$ for $Gr_{ac} = 10^3$ is 64 \% larger than for $Gr_{ac} = 10^4$ (not shown), resulting in a 128 \% difference on the viscous force estimate. Since the estimate of the velocity peak position is obtained by balancing the acoustic with the viscous forces, the error on $R_{jet}$ thus causes $x_{p_v}$ to diverge away from the numerical simulation results for $Gr_{ac} < 10^4$. Thus, considering the same $R_{jet}$ for all $Gr_{ac}$ in Eq.~\eqref{eq:inertial_regime_balance_viscosity} leads to an overestimation of the viscous forces at low $Gr_{ac}$.

The same loss of accuracy at low $Gr_{ac}$ is found for $U_{x,p_v}$, albeit to a lesser extent (Fig.~\ref{fig:velocity_peak_position_and_amplitude}~(b)). Although $U_{x,p_i}$ provides an acceptable discrepancy of 8.7~\% with the velocity peak magnitudes observed in the numerical simulations, its variation with $Gr_{ac}$ is further improved when viscous forces are taken into account upstream the peak. The difference between $U_{x,p_v}$ and $u_{x,p}$ indeed falls below 3~\% for $Gr_{ac}^{3/5} / S^{2/5} \geq 4 \times 10^2$. For comparison, we added the experimental data points of \citet{Dentry2014} (empty symbols) to Fig.~\ref{fig:velocity_peak_position_and_amplitude}~(b). In their experiments, the authors used rectangular Rayleigh Surface Acoustic Waves (SAW) devices to drive streaming jets. Contrary to our numerical setup, the jets in Ref.~\citep{Dentry2014} were forced by a beam of rectangular cross section and were therefore non-axisymmetric. These differences in the forcing and flow geometries inevitably affect the magnitudes of the forces to which the jet is subjected, causing the values of $u_{x,p}$ to differ. Nevertheless, the data points of Ref.~\citep{Dentry2014} scale similarly with $Gr_{ac}$ and $S$: this proves that $u_{x,p}$ indeed satisfies a local balance between the acoustic and viscous forces, as assumed by the scaling laws~\eqref{eq:velocity_peak_amplitude_inviscid} and \eqref{eq:velocity_peak_amplitude}. Improving the match between these scaling laws and the $u_{x,p}$ of Ref.~\citep{Dentry2014} would require developments specific to their peculiar jet geometry that are out of the scope of the present work.

\subsection{Viscosity-dominated flow region downstream the velocity peak} \label{subsection:viscous_regime}

We shall finally consider the fluid region located downstream $x_p$, where the on-axis jet velocity decays smoothly before reaching the wall at $x = L$ (blue region in the bottom plot of Fig.~\ref{fig:velocityField_GrAc5000}). Since inertial effects become negligible near $x_p$, the on-axis velocity $U_{x,d}$ in the downstream region results from the balance between the acoustic and the viscous forces, so using the definition of the acoustic power (Eq.~\eqref{eq:OM_intensity}):
\begin{equation} \label{eq:viscous_regime}
\frac{U_{x,d}}{R_{jet}^2} \sim \frac{ Gr_{ac} }{16 R_{beam}^2} e^{- \frac{x}{L_{\alpha}} } \, .
\end{equation}
In Ref.~\cite{Moudjed2014} a similar longitudinal rate of change is assumed for $R_{jet}$ and $R_{beam}$, 
so the variations of $U_{x,d}$ could only arise from the attenuation of the beam. 
This is not observed here (Fig.~\ref{fig:attenuation_effects_velocity_radii_profiles}~(b)), and beam attenuation is too weak to be the sole reason for the decay of $u_x$ downstream $x_p$ (Fig.~\ref{fig:velocityField_GrAc5000}).
\begin{figure}[b]
	\centering
	\includegraphics[width=0.98\textwidth]{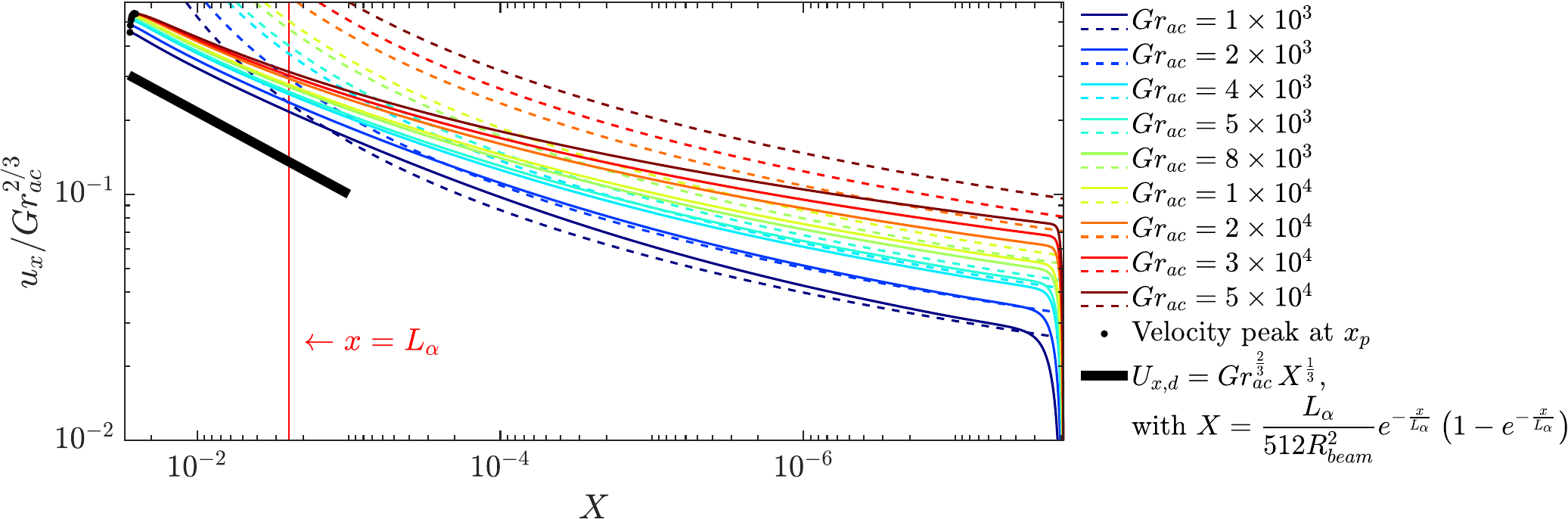}
	\caption{On-axis velocity profiles downstream the velocity peaks $x_p$, which are represented by the black points. THe profiles are shown for different $Gr_{ac}$. The quantity plotted on the $x$-axis includes the longitudinal variations of the acoustic force caused by both diffraction and attenuation of the beam (see the right-hand side of Eq.~\eqref{eq:OM_viscous}). The red vertical line locates the force attenuation distance $L_{\alpha} = L / \left( 2 N \right)$. The scaling law~\eqref{eq:OM_viscous} for the on-axis velocity $U_{x,d}$ at $x_p \leq x \lessapprox L_{\alpha}$ is represented by the thick straight black line. The colored solid lines correspond to the numerical data. The dashed curves represent the self-similar jet solution $U_{x,S}$ of Schlichting~\citep{Schlichting2000} given by Eq.~\eqref{eq:Schlichting_self-similar_solution} and based on the maximum axial momentum flux $M = \pi L Gr_{ac} / \left( 32 N \right) = \pi L_{\alpha} Gr_{ac} / 16$ asymptotically acquired by the jet.}
	\label{fig:viscous_regime}
\end{figure}
It is thus of primary importance to rely on a good estimate of the axial jet profile to assess viscosity effects. Contrary to Ref.~\citep{Moudjed2014}, we do not assume $R_{jet} = R_{beam}$, but instead base $R_{jet}$ on equation~\eqref{eq:OM_axial_momentum} describing the build up of the jet momentum flux along $x$. Equation~\eqref{eq:viscous_regime} then gives:
\begin{equation} \label{eq:OM_viscous}
U_{x,d} \sim Gr_{ac}^{ \frac{2}{3} } X^{ \frac{1}{3} } \, , \quad \text{with} \ X = \frac{L_{\alpha}}{512 R_{beam}^2} e^{ - \frac{x}{L_{\alpha}} } \left( 1 - e^{- \frac{x}{L_{\alpha}} } \right) \, .
\end{equation}
As the considered flow region is located downstream $L_F$, the enlargement of the beam due to diffraction is significant and needs to be taken into account in the definition of $X$. This means that $R_{beam}$ can no longer be considered as constant, but shall instead follow the linear expansion given by Eq.~\eqref{eq:beam_radius}. Thus, from Eq.~\eqref{eq:OM_viscous}, the longitudinal variations of $u_x$ are caused by both diffraction and attenuation.

As shown in Fig.~\ref{fig:viscous_regime}, $U_{x,d}$ (plotted as a thick straight black line) correctly captures the slope the on-axis $u_x$ (solid colored curves) on $x_p \leq x \leq L_{\alpha}$ and for all $Gr_{ac}$. Besides, $U_{x,d}$ differs from $u_x$ by a factor at most 3, meaning that Eq.~\eqref{eq:OM_viscous} also correctly estimates the order of magnitude of $u_x$ on that $x$ interval.

\subsection{Negligible acoustic force region far downstream the velocity peak} \label{subsection:negligible_force_regime}

Beyond $L_{\alpha}$, the magnitude of the acoustic force relative to inertia decreases due to attenuation, causing $u_x$ to deviate from $U_{x,d}$. Equation~\eqref{eq:OM_viscous} is thus unable to describe the on-axis $u_x$ over the entire $x \geq x_p$ jet region, and another approach is needed to recover the on-axis profile of $u_x$ at $x \gg L_{\alpha}$.

At $x \gg L_{\alpha}$, most of the energy of the acoustic beam has been attenuated, meaning that the forcing is negligible whatever the $Gr_{ac}$. As the pressure forces also are negligible, the jet necessarily satisfies a balance between inertial and viscous forces at $x \gg L_{\alpha}$ according to the Navier-Stokes equations~\eqref{eq:Navier-Stokes_dimensionless}. Furthermore, as shown in Sec.~\ref{subsection:attenuation_effects}, the beam attenuation causes the axial momentum of the acoustic streaming jet to asymptotically reach the constant value $M = \pi L Gr_{ac} / \left( 32 N \right) = \pi L_{\alpha} Gr_{ac} / 16 $. The situation thus becomes similar to the case of a straight jet of constant axial momentum flux $M$ emerging from an orifice and described by the boundary layer equation:
\begin{equation} \label{eq:boundary_layer}
u_x  \frac{\partial u_x}{\partial x} + u_r \frac{\partial u_x}{\partial r}= \frac{1}{r} \frac{\partial  }{\partial r} \left( r \frac{\partial u_x}{\partial r} \right) \, ,
\end{equation}
for which a divergence-free self-similar solution is~\cite{Schlichting2000}:
\begin{equation*}
u_x = \frac{3}{8 \pi} \frac{M}{x} \frac{1}{\left( 1 + \zeta^2 \right)^2} \, ,
\end{equation*}
where
\begin{equation*}
\zeta = \frac{1}{8} \sqrt{ \frac{3 M}{\pi} } \frac{r}{x}
\end{equation*}
is the self-similar variable. The on-axis velocity of the streaming jet is then:
\begin{equation} \label{eq:Schlichting_self-similar_solution}
U_{x,S} = \frac{3 L_{\alpha} Gr_{ac}}{128x} \, ,
\end{equation}
which is thus expected to accurately describe the on-axis jet velocity as $x \rightarrow +\infty$.

The scaling law~\eqref{eq:Schlichting_self-similar_solution} for $U_{x,S}$ is plotted as dashed curves for all $Gr_{ac}$ in Fig.~\ref{fig:viscous_regime} alongside the numerical velocity profiles (solid curves). As expected, the decrease of the acoustic forcing relative to inertia as $x$ increases (or equivalently, as $X$ reduces) causes the numerical profiles to come closer to $U_{x,S}$. This strong agreement confirms the suitability of $U_{x,S}$ for describing the on-axis jet velocity as $x \rightarrow +\infty$.

\subsection{Global picture of the streaming velocity profile on the jet axis}
\label{subsection:summary_scaling_laws}

\begin{figure}[h!]
	\centering
	\includegraphics[trim={1cm, 0.25cm, 1cm, 1cm}, clip, scale=0.5]{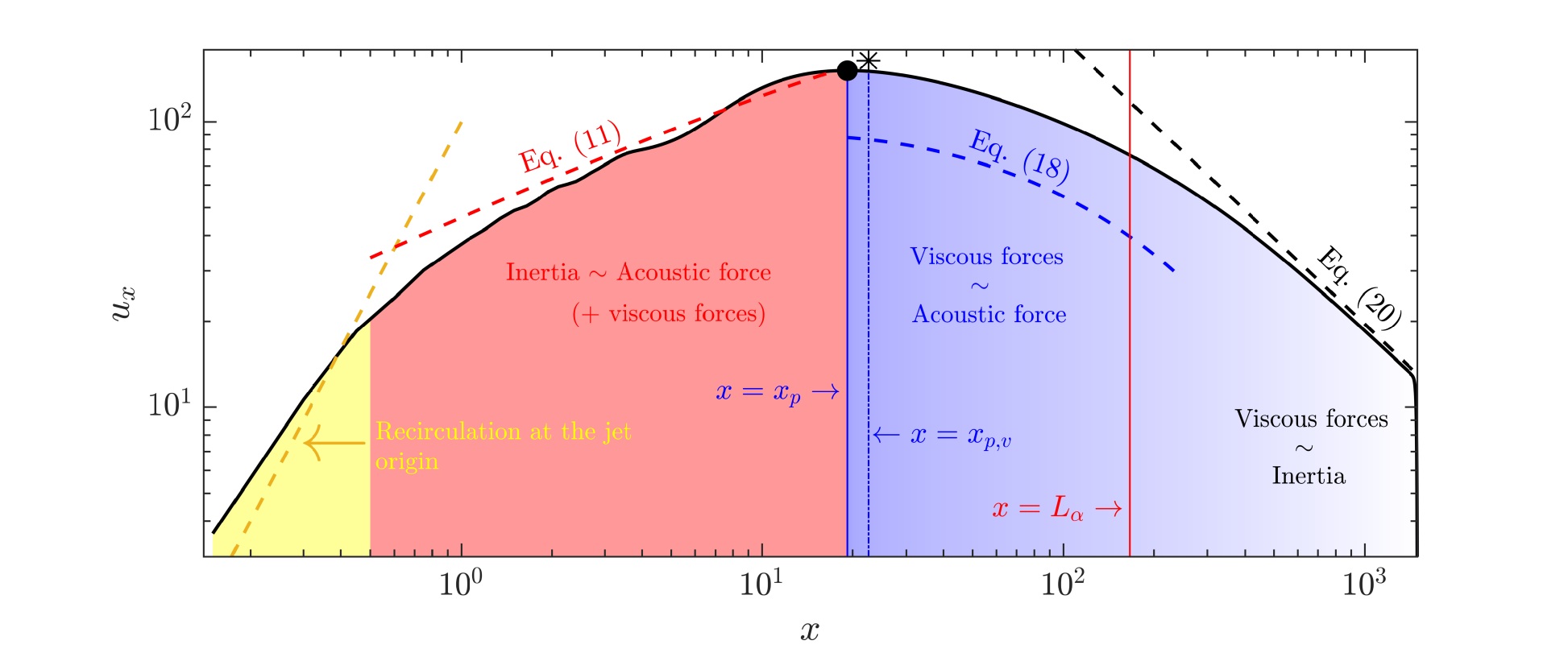}
	\caption{Longitudinal profile of the on-axis axial velocity $u_x$ computed for $Gr_{ac} = 5 \times 10^3$ (solid black line), together with the scaling laws (dashed lines) and the force balance in each jet region. The scaling law in the yellow region is taken from \citet{Moudjed2015}. The predicted peak position $x_{p_v}$ (Eq.~\eqref{eq:velocity_peak_position_viscous}) and amplitude $U_{x,p_v}$ (from Eq.~\eqref{eq:velocity_peak_amplitude} with $x_{p_v}$) are represented by the dash-dotted blue vertical line and the black star, respectively. The solid blue and red vertical lines  respectively locate the actual on-axis velocity peak $x_p$ and the force attenuation distance $L_{\alpha} = L / \left( 2 N \right)$.}
	\label{fig:all_scaling_laws_together}
\end{figure}

All the scaling laws derived throughout this work for a steady and axisymmetric Eckart streaming jet are summarized in Fig.~\ref{fig:all_scaling_laws_together} in the $Gr_{ac} = 5 \times 10^3$ case. The scaling law given in Ref.~\cite{Moudjed2015} and describing the jet formation at $x \leq 0.5$ has also been plotted.
Downstream this small region dominated by 3D effects, our quasi-one dimensional approach sheds lights on the following successive regions:
\setlist{nolistsep}
\begin{enumerate}[label=(\roman*), noitemsep]
	\item First, the acoustic force strongly accelerates the fluid near the source (red region in Fig.~\ref{fig:all_scaling_laws_together}). In this region of the jet, the acoustic forcing is essentially balanced by inertial forces \citep{Mitome1995,Moudjed2014,Orosco2022}. However, for the investigated values of $Gr_{ac}$, viscous forces are not entirely negligible, causing a discrepancy of nearly 50~\% between the simulated profiles and the scaling law of \citet{Moudjed2014}. Accounting for the viscous friction forces leads to Eq.~\eqref{eq:OM_inertial_corrected}, as proposed in Sec.~\ref{subsection:inertial_regime}. This new scaling law clearly improves the velocity scaling on $Gr_{ac}$ and reduces to 25~\% the error on the predicted velocity. Finally, the oscillations of $u_x$ observed in Fig.~\ref{fig:transducer_mesh_convergence}(b) are directly correlated to the strong longitudinal gradients of the acoustic force in the near-acoustic field~\cite{Moudjed2015}.
	\item The acceleration region is followed by a velocity peak on the jet axis. Beam diffraction causes the acoustic force on the axis to significantly decrease locally, so that the acceleration becomes weaker, up to the point where inertia completely vanishes at the velocity peak position $x_p$. By balancing the acoustic and viscous forces, we obtained a scaling law for $x_p$ and the peak velocity magnitude $u_{x,p}$. The corresponding star on Fig.~\ref{fig:all_scaling_laws_together} has been plotted using the derived laws given by Eqs.~\eqref{eq:velocity_peak_position_viscous} and~\eqref{eq:velocity_peak_amplitude}. These two relations clearly allow to tailor streaming jets to e.g. maximize the velocity at a given point in the fluid domain by tuning $S$ or $Gr_{ac}$ for instance.
	\item Further downstream, as the beam keeps opening up due to diffraction, the acoustic force becomes locally weaker and is balanced by viscous forces. As described by Eq.~\eqref{eq:OM_viscous}, the beam diffraction and attenuation jointly drive the smooth velocity decay along the jet axis for $x_p \leq x \lesssim L_{\alpha}$ (deep blue region of Fig.~\ref{fig:all_scaling_laws_together}). 
	\item For $x \gg L_{\alpha}$, the energy of the acoustic beam is almost entirely attenuated. Thus, the momentum balance reduces to an equilibrium between viscous and inertial forces. Besides, the axial momentum of the jet saturates to the constant value $M = \pi L Gr_{ac} / \left( 32 N \right)$ (from Eq.~\eqref{eq:axial_momentum_theoretical}). As a consequence, the scaling law of the jet recovers that of a self-similar axisymmetric jet of constant momentum~(Eq.\eqref{eq:Schlichting_self-similar_solution}) adapted from~\citep{Schlichting2000} and represented by the dashed black line in Fig.~\ref{fig:all_scaling_laws_together}.
\end{enumerate}

%%%%%%%%%%%%%%%%%%%%%%%%%%%%%%%%%%%%%%%%%%%%%%%%%%%%%%%%%%%%%%%%%%%%%%%%%%%%%%%%%%%%%%%%%%
%%%%%%%%%%%%%%%%%%%%%%%%%%%%%%%%%%%%%%%%%%%%%%%%%%%%%%%%%%%%%%%%%%%%%%%%%%%%%%%%%%%%%%%%%%
%%%%%%%%%%%%%%%%%%%%%%%%%%%%%%%%%%%%%%%%%%%%%%%%%%%%%%%%%%%%%%%%%%%%%%%%%%%%%%%%%%%%%%%%%%
%%%%%%%%%%%%%%%%%%%%%%%%%%%%%%%%%%%%%%%%%%%%%%%%%%%%%%%%%%%%%%%%%%%%%%%%%%%%%%%%%%%%%%%%%%

\section{Conclusion} \label{section:conclusion}

We presented a numerical study of an Eckart-type acoustic streaming jet representative of former experiments~\citep{Kamakura1996,Mitome1998,Frenkel2001,Moudjed2014}. The flow develops in a wide and elongated cylindrical cavity entirely filled with a viscous liquid. The acoustic source is a plane circular transducer, of axis aligned with the cylinder axis, placed at one end of the cavity in direct contact with the fluid. It radiates an axisymmetric beam-shaped acoustic field yielding a body force that drives the flow. An important aspect of this work is that the characteristic attenuation length $L_{\alpha}$ of the  acoustic force is smaller but comparable to the cavity length, so that the jet features both a region close to the source with prominent acoustic force,  and regions further away where it is not. The diameter of the cavity is sufficiently large to avoid the influence of the lateral wall on the jet. The range of acoustic force magnitudes, measured by the acoustic Grashof number $Gr_{ac}$, is limited to moderate values, so that the resulting streaming jet remains axisymmetric, steady and laminar. Hence, this configuration provides a generic representation of applications where a transducer is directed into a large volume so as to act over a distance greater than the attenuation length but smaller than the size of the cavity (this would prove particularly useful in metallurgical and crystal growth applications~\citep{Kozhemyakin1998,Miralles2023}). In this type of configuration, acoustic attenuation plays a central role for the modeling of the acoustic field even at small distance from the source. Indeed, the error incurred on the injected momentum flux when neglecting attenuation is already 10~\% at a distance $0.2 L_{\alpha}$ from the source, and further increases with the on-axis distance $x$ from the transducer. Despite being frequently used in the literature, the unattenuated beam approximation leads to significant errors: it severely overestimates the jet velocity along its axis, which in turn affects the flow structure by reducing the impact of viscous diffusion on the radial spreading of the jet.

To characterize the jet, several flow regions and features were identified based on the mechanisms driving the flow. For each of these regions, we derived scaling laws describing the on-axis velocity profile. The obtained relations are all summarized in Fig.~\ref{fig:all_scaling_laws_together} and section~\ref{subsection:summary_scaling_laws}. We thus give a complete picture of the different regimes encountered along an acoustic streaming jet that is several times longer than the attenuation length. These scaling laws offer an advantageous alternative to simulations for obtaining an accurate representation of the on-axis jet velocity profile. This paves the way for tailored applications informed by accurate predictions on acoustically-driven jets. For instance, we may use these scaling laws to optimize configurations in which several successive beam reflections are used to create a fully 3D forcing~\citep{Vincent2024}. Matching the velocity peak position $x_p$ with a beam reflection on a wall would then maximize the velocity at the corresponding jet impingement, hence maximizing the 3D flow intensity and mixing efficiency~\citep{Vincent2024}. A similar choice may also be made to maximize the shear stresses at the impingement to e.g. improve the mass transfer at a wall~\citep{ElGhani2021}, or for surface cleaning~\citep{Busnaina1998}. The jet can also be optimized to favor transition to a chaotic or turbulent state for the purpose of increasing mixing by contactless means~\citep{Launay2019}.

Finally, it would be interesting to see whether the different approaches used in the present work could be applied to the study of other flows forced by external beam-shaped fields. Despite obvious differences related to the type of waves driving the flow, Eckart streaming indeed shares common aspects with flows driven by the attenuation of laser light~\cite{Savchenko1997,Shneider2016,Wunenburger2011} or internal gravity wave beams~\cite{Bordes2012,Grisouard2013,Dauxois2018}. These similarities may potentially make it possible to transpose some of the derivations from this paper to such flows.

\begin{acknowledgments}
This work was carried out as part of the BRASSOA project supported by the Institut Carnot Ingénierie@Lyon and by a Royal Society International Exchange grant (Ref.~IES\textbackslash R2\textbackslash 202212). The support from the PMCS2I of Ecole Centrale de Lyon for the numerical  calculations is gratefully acknowledged. We would particularly like to thank both Laurent Pouilloux (Ecole Centrale de Lyon) and Alex Pedcenko (Coventry University) for their availability and for providing help at any stage of this project. We also would like to thank Hugh M. Blackburn for helping with the use of the spectral element code Semtex.

For the purpose of Open Access, a CC-BY public copyright licence
has been applied by the authors to the present document and will
be applied to all subsequent versions up to the Author Accepted Manuscript arising from this submission.
\end{acknowledgments}

\appendix
\section{Derivation of the equations for the acoustic field radiated by a plane circular transducer}
\label{appendix:acoustic_force_derivation}

We shall derive the equations governing the attenuated acoustic field radiated by a plane circular transducer. For the sake of clarity, the derivation is carried out using dimensional equations. We discuss at the end of the appendix the nondimensionalization and normalization of the forcing field equations using the same length, mass and time scales as those of the hydrodynamics problem. We restrict ourselves to the case of linear acoustic propagation but include dissipation.

\subsection{Governing equations of acoustics}

Although acoustic streaming refers to an incompressible flow, it relies on the attenuation of sound waves, which are a compressible process~\cite{Chassaing2002}. Based on the assumptions listed in Sec.~\ref{subsection:acoustic_field}, the propagation of linear sound waves far from solid boundaries can be described by the irrotational compressible Navier-Stokes and continuity equations linearized around a resting fluid of density $\rho$, together with an isentropic equation of state~\citep{Kinsler2000,Blackstock2000}:
\begin{subequations}
\begin{align}
&\rho \frac{\partial \bm{u_{ac}} }{\partial t} = - \nabla p_{ac} + \rho c^2 \tau \bm{\nabla} \left( \bm{\nabla} \cdot \bm{u_{ac} } \right) \, , \label{eq:compressible_Navier-Stokes} \\
&\frac{ \partial \rho_{ac} }{\partial t} + \rho \bm{\nabla} \cdot \bm{u_{ac} } = 0 \, , \label{eq:compressible_continuity}\\
&p_{ac} = c^2 \rho_{ac} \, , \label{eq:equation_of_state}
\end{align}
\end{subequations}
where $p_{ac}$ and $\rho_{ac}$ are the acoustic fluctuations of pressure and density, respectively, $c$ is the speed of sound and $\displaystyle{\tau = \frac{1}{\rho c^2} \left( \frac{4}{3} \mu + \eta \right)}$. The propagation of linear attenuated sound waves is then governed by a wave equation for $\bm{u_{ac}}$:
\begin{equation*}
\frac{1}{c^2} \frac{ \partial^2 \bm{u_{ac}} }{\partial t^2} - \bm{\nabla}^2 \bm{u_{ac}} = \tau \frac{ \partial }{\partial t} \left( \bm{\nabla}^2 \bm{u_{ac}} \right) \, ,
\end{equation*}
which is obtained by (i) taking the time-derivative of Eq.~\eqref{eq:compressible_Navier-Stokes}, (ii) taking the gradient of Eq.~\eqref{eq:compressible_continuity}, (iii) combining the obtained equations and (iv) making an extensive use of $\bm{\nabla} \times \bm u_{ac} = \bm{0}$. For a monochromatic wave $\bm{u_{ac}} = \bm{\widehat{u}_{ac}} e^{j \omega t}$, the above wave equation then reduces to a complex Helmholtz equation:
\begin{equation} \label{eq:HelmholtzEquation}
\bm{\nabla}^2 \bm{\widehat{u}_{ac}} + k^2 \bm{\widehat{u}_{ac} } = \bm{0} \, .
\end{equation}
In Eq.~\eqref{eq:HelmholtzEquation}, the effects of attenuation appear through the imaginary part of the complex wavenumber $k$ defined as:
\begin{equation} \label{eq:wavenumber_definition}
k = k_a - j \alpha = \frac{ \omega }{c \sqrt{ 1 + j \omega \tau }} \, ,
\end{equation}
where $j^2 = -1$ and  $\omega \tau$ is a dimensionless group comparing the wavelength to the sound attenuation length (i.e., $\omega \tau = N S / \left( 1.22 L \right)$. The real and imaginary parts of $k$ are given by~\citep{Kinsler2000}:

\begin{eqnarray} \label{eq:wavenumber_real_imaginary_parts}
%\begin{aligned}[t]
k_a &= \frac{ \omega }{c \sqrt{2} } \left[ \frac{ \sqrt{1 + \left( \omega \tau \right)^2 } + 1 }{ 1 + \left( \omega \tau \right)^2 } \right]^{\frac{1}{2}}
&\underset{\omega \tau \ll 1}{\sim} \frac{\omega}{c} \, \\
%\end{aligned}
%\hspace{3cm}
%\begin{aligned}[t]
\alpha &= \frac{ \omega }{c \sqrt{2}} \left[ \frac{ \sqrt{ 1 + \left( \omega \tau \right)^2 } - 1 }{ 1 + \left( \omega \tau \right)^2 } \right]^{ \frac{1}{2} } 
&\underset{\omega \tau \ll 1}{\sim} \frac{ \omega^2 \tau }{2 c} = \frac{\omega^2}{2 \rho c^3} \left( \frac{4}{3} \mu + \eta \right) \, .
%\end{aligned}
\end{eqnarray}

Finally, under the assumptions (ii) and (iii) of Sec.~\ref{subsection:acoustic_field}, the momentum equation~\eqref{eq:compressible_Navier-Stokes} can be rewritten to give:
\begin{equation} \label{eq:generalized_acoustic_impedance}
j \frac{\rho c^2 k^2}{\omega} \bm{\widehat{u}_{ac}} = - \nabla \widehat{p}_{ac} \, ,
\end{equation}
which can be seen as a generalized specific acoustic impedance. Equation~\eqref{eq:generalized_acoustic_impedance} proves particularly useful when computing the acoustic intensity $\bm{I}$:
\begin{equation} \label{eq:acoustic_intensity_definition}
\bm{I} = \frac{1}{2} \Re \left\{ p_{ac} \bm{u_{ac}}^* \right\} \, ,
\end{equation}
where $^*$ is the complex conjugate.

\subsection{Integral relations for the acoustic intensity field radiated by a plane circular transducer}

\begin{figure}[t!]
	\centering
	\includegraphics[trim={0, 3cm, 0, 2cm}, clip, width=0.6\textwidth]{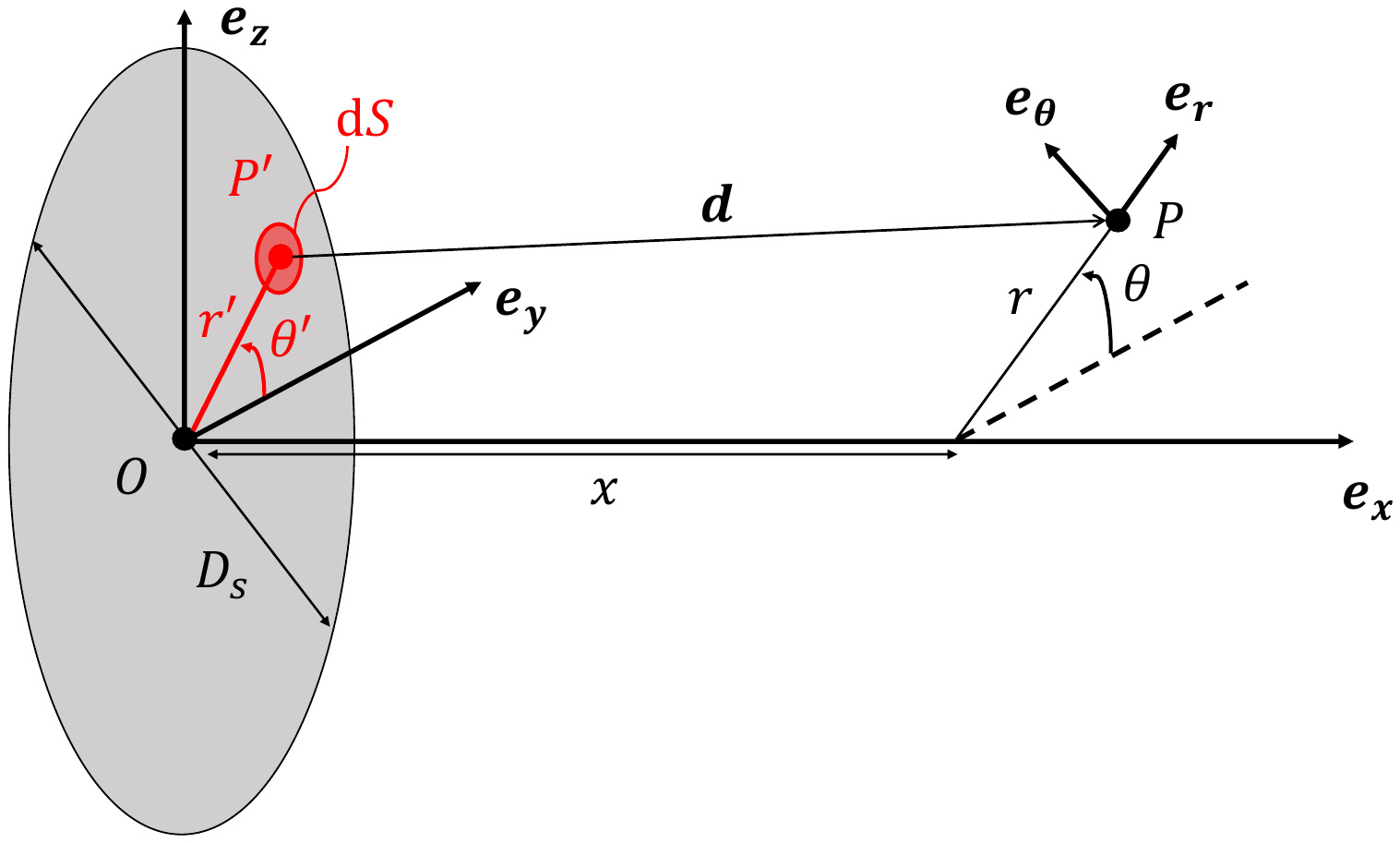}
	\caption{Sketch of the circular transducer of diameter $D_s$ (gray), modeled as a plane disc made of an infinite number of point sources. Each source is centered on an elementary surface $\mathrm{d}S$ and radiates spherical acoustic waves contributing to the net acoustic pressure and velocity (hence intensity) levels at $P$ in the fluid domain.}
	\label{fig:transducer_sketch}
\end{figure}

To obtain the integral expression for $p_{ac}$ and $\bm{u_{ac}}$ (and hence to infer $\bm{I}$), we shall closely follow the derivation of Blackstock~\citep{Blackstock2000} and adapt it to account for sound attenuation. The idea is to model the vibrating surface $S_t$ of the transducer as a plane disc made of an infinite number of point sources. Each of these sources is centered on an elementary surface $\mathrm{d}S$ (Fig.~\ref{fig:transducer_sketch}) and drives a net flow rate $Q_0$ through that surface. We shall thus first focus on the spherical pressure wave emitted by a single point source, and determine how the wave amplitude relates to $Q_0$. The acoustic pressure and velocity levels in the fluid domain are then obtained by summing up the spherical waves radiated by all point sources on $S_t$. We checked that the same pressure Rayleigh integral as Eq.~\eqref{eq:lossy_Rayleigh_integral_pressure} can be obtained using Green's functions (see Ref.~\citep{Pierce1981} for the derivation in the context of unattenuated acoustics).

A point source may be seen as a special case of a pulsating sphere of vanishing mean radius $R_S$.
As its surface vibrates at a velocity $U_{R_S} e^{j \omega t} \bm{e_{d'}}$, where $\bm{e_{d'}}$ is the unit radial vector, the sphere produces spherical acoustic pressure waves defined as~\citep{Blackstock2000}:
\begin{equation} \label{eq:spherical_pressure_wave}
p_{ac} = \frac{A}{d'} e^{j \left[ \omega t - k \left( d' - R_S \right) \right]} \, ,
\end{equation}
with $d'$ the distance from the center of the sphere, and $A$ is a constant defining the initial wave amplitude. The pulsating sphere drives a volumetric flow rate of amplitude $Q_{R_S} = 4 \pi R_S^2 U_{R_S}$ through its surface. From Eqs.~\eqref{eq:spherical_pressure_wave} and \eqref{eq:generalized_acoustic_impedance}, one obtains:
\begin{equation*}
Q_{R_S} = -j \frac{4 \pi A \omega \left( 1 + j k R_S \right) }{\rho c^2 k^2} \, .
\end{equation*}
The strength $Q_0$ of a point source is then determined by taking the limit of $Q_{R_S}$ as $R_S \rightarrow 0$, allowing to relate $Q_0$ to $A$. Equation~\eqref{eq:spherical_pressure_wave} can thus be rewritten as:

\begin{equation} \label{eq:spherical_pressure_wave_defined_amplitude}
p_{ac} = j \frac{\rho c^2 k^2 Q_0}{4 \pi \omega d'} e^{j \left( \omega t - k d' \right) } \, .
\end{equation}
On $S_t$, $Q_0 = \bm{u_0} \cdot \bm{n_t} \mathrm{d}S$, with $\bm{u_0}$ the vibrating velocity of the transducer and $\bm{n_t}$ the normal unit vector. Because the superposition principle applies, $p_{ac}$ in the fluid domain is a linear combination of all the pressure waves emitted by each point source and is thus obtained by integrating Eq.~\eqref{eq:spherical_pressure_wave_defined_amplitude} over $S_t$:
\begin{equation} \label{eq:lossy_Rayleigh_integral_pressure}
p_{ac} = j \frac{ \rho c^2 k^2 u_0 }{2 \pi \omega} \iint_{S_t} \frac{ e^{j \left( \omega t - k \Vert \bm{d} \Vert \right)} }{ \Vert \bm{d} \Vert } \mathrm{d}S \, ,
\end{equation}
in which $Q_0$ has been doubled, since each point source radiates waves in the $x >0$ semi-infinite domain only~\citep{Blackstock2000}. Note that the same result can be more rigorously obtained using the formalism of Green's functions: the factor 2 then results from the method of images applied at the transducer surface at $x=0$ (see the full derivation in \citet{MorseFeshbach} for unattenuated acoustics). As the transducer vibrates uniformly, the acoustic fields are axisymmetric and thus need to be calculated in a single constant-$\theta$ half-plane only. For simplicity, we shall consider $\theta = 0$, so that the vectorial distance $\bm{d}$ between a point source at $\left( r', \theta' \right)$ on $S_t$ and the point at $\left( x, r \right)$ where $p_{ac}$ is evaluated (Fig.~\ref{fig:transducer_sketch}) reduces to:
\begin{equation*}
\bm{d} = x \, \bm{e_x} + \left[ r - r' \cos \left( \theta' \right) \right] \, \bm{e_r} - r' \sin \left( \theta' \right) \bm{e_{\theta}} \, .
\end{equation*}

An integral expression for $\bm{u_{ac}}$ is obtained using a similar approach. From Eqs.~\eqref{eq:generalized_acoustic_impedance} and \eqref{eq:spherical_pressure_wave_defined_amplitude}, the acoustic velocity induced by a single point source is
\begin{equation*}
\bm{u_{ac}} = \frac{Q_0}{4 \pi d'^2} \left( 1 + j k d' \right) e^{ j \left( \omega t - k d' \right) } \, \bm{e_{d'}} \, .
\end{equation*}
The acoustic velocity radiated by the transducer is then obtained by integrating over $S_t$:
\begin{equation} \label{eq:lossy_Rayleigh_integral_velocity}
\bm{u_{ac}} = \frac{ u_0 }{2 \pi} \iint_{S_t} \frac{ \left( 1 + j k \Vert \bm{d} \Vert \right) }{ \Vert \bm{d} \Vert^2 } \frac{ \bm{d} }{ \Vert \bm{d} \Vert } e^{ j \left( \omega t - k \Vert \bm{d} \Vert \right) } \mathrm{d}S \, ,
\end{equation}
in which $Q_0$ has again been doubled for the same reason as in Eq.~\eqref{eq:lossy_Rayleigh_integral_pressure}. We stress that our formalism easily yields integral relations for both $p_{ac}$ and $\bm{u_{ac}}$, making the numerical implementation quite efficient and independent of the grid used for flow computations. Green's function formalism on the other hand, cannot give a straightforward derivation of $\bm{u_{ac}}$ as in Eq~\eqref{eq:lossy_Rayleigh_integral_velocity}. From Eq.~\eqref{eq:acoustic_intensity_definition}, it appears that the temporal variations of $p_{ac}$ and $\bm{u_{ac}}$ do not affect $\bm{I}$; we can thus drop $e^{j \omega t}$ from Eqs.~\eqref{eq:lossy_Rayleigh_integral_pressure} and \eqref{eq:lossy_Rayleigh_integral_velocity}.

\subsection{Analytical expressions of the acoustic fields on the transducer axis}

Except at particular points in the fluid domain, the integral expressions for $p_{ac}$ and $\bm{u_{ac}}$ derived in the previous section cannot be computed analytically in the entire $x > 0$ domain, and shall thus be evaluated numerically. It is nevertheless possible to obtain analytical expressions for $p_{ac}$, $\bm{u_{ac}}$ and $\bm{I}$ on the acoustic axis; these expressions prove useful for checking the numerical computations of Eqs.~\eqref{eq:lossy_Rayleigh_integral_pressure} and \eqref{eq:lossy_Rayleigh_integral_velocity}.

We shall first evaluate $p_{ac}$ at $r=0$, where the attenuated Rayleigh integral~\eqref{eq:lossy_Rayleigh_integral_pressure} is:
\begin{equation*}
p_{ac}(x, r=0) = j \frac{ \rho c^2 k^2 u_0 }{2 \pi \omega} e^{j \omega t}  \int_{\theta'=0}^{2\pi} \int_{r'=0}^{D_s / 2} \frac{ e^{- j k \sqrt{ x^2 + r'^2 } } }{ \sqrt{x^2 + r'^2} } \, r' \, \mathrm{d}r' \, \mathrm{d}\theta' \, .
\end{equation*}
Setting $X = \sqrt{x^2 + r'^2}$, the above integral can be computed analytically to give:
\begin{equation*}
\begin{split}
p_{ac}(x, r=0) &= \frac{ \rho c^2 k u_0 }{\omega} e^{j \omega t} \left( e^{- j k x} - e^{- j k \sqrt{ x^2 + \frac{D_s^2}{4} } } \right) \\
&= \frac{ \rho c^2 k u_0 }{\omega} e^{j \omega t} e^{-j \frac{k}{2} \left( \sqrt{ x^2 + \frac{D_s^2}{4} } + x \right) } \left[ e^{ j \frac{k}{2} \left( \sqrt{ x^2 + \frac{D_s^2}{4} } - x \right) } - e^{- j \frac{k}{2} \left( \sqrt{ x^2 + \frac{D_s^2}{4} } - x \right) } \right] \\
&= j \frac{2 \rho c^2 k u_0 }{\omega} e^{j \omega t} e^{ - j \frac{k}{2} \left( \sqrt{ x^2 + \frac{D_s^2}{4} } + x \right) } \, \sin \left[ \frac{k}{2} \left( \sqrt{ x^2 + \frac{D_s^2}{4} } - x \right) \right] \, .
\end{split}
\end{equation*}
Then, substituting the first equality of Eq.~\eqref{eq:wavenumber_definition} for $k$ in the above equation finally gives:
\begin{equation} \label{eq:analytical_pressure_acoustic_axis}
\begin{split}
p_{ac}(x,r=0) =& \frac{2 \rho c^2 \left( k_a - j \alpha \right) u_0}{\omega} e^{ - \frac{\alpha}{2} \left( \sqrt{x^2 + \frac{D_s^2}{4} } + x \right)} e^{ j \left[ \omega t - \frac{k_a}{2} \left( \sqrt{x^2 + \frac{D_s^2}{4} } + x \right) \right] } \\
&\times \Bigg( \sinh \left[ \frac{\alpha}{2} \left( \sqrt{x^2 + \frac{D_s^2}{4} } - x \right)  \right] \cos \left[ \frac{k_a}{2} \left( \sqrt{x^2 + \frac{D_s^2}{4} } - x \right)  \right] \\
&+ j \sin \left[ \frac{k_a}{2} \left( \sqrt{x^2 + \frac{D_s^2}{4} } - x \right)  \right] \cosh \left[ \frac{\alpha}{2} \left( \sqrt{x^2 + \frac{D_s^2}{4} } - x \right)  \right] \Bigg) \, .
\end{split}
\end{equation}
Equation~\eqref{eq:analytical_pressure_acoustic_axis} is an original result accounting for attenuation. To assess its validity, we checked that taking the modulus of Eq.~\eqref{eq:analytical_pressure_acoustic_axis} with $\alpha = 0$ gives the same unattenuated $\vert p_{ac} \vert$ as in Ref.~\citep{Kinsler2000}. 

Contrary to $p_{ac}$, Eq.~\eqref{eq:lossy_Rayleigh_integral_velocity} cannot be calculated analytically even for $r=0$. However, $\bm{u_{ac}}(x, r=0) = u_{ac,x}(x, r=0) \bm{e_x}$ can be computed using Eq.~\eqref{eq:generalized_acoustic_impedance}, in which $\nabla p_{ac}$ is calculated from Eq.~\eqref{eq:analytical_pressure_acoustic_axis}. The resulting expression of $u_{ac,x}$ can finally be used with Eq.~\eqref{eq:acoustic_intensity_definition} to get the on-axis acoustic intensity $I_x(x, r=0)$:
\begin{equation} \label{eq:analytical_intensity_acoustic_axis}
\begin{split}
I_x(x, r=0) =& \frac{\rho c^2 u_0^2 k_a }{2 \omega} e^{ - \alpha \left( \sqrt{x^2 + \frac{D_s^2}{4} } + x \right)} \Bigg( \left[ 1 - \frac{x}{ \sqrt{ x^2 + \frac{D_s^2}{4} } } \right] \sinh \left[ \alpha \left( \sqrt{ x^2 + \frac{D_s^2}{4} } - x \right) \right] \\
&+ \left[ \frac{x}{ \sqrt{x^2 + \frac{D_s^2}{4}} } + 1 \right] \left( \cosh \left[ \alpha \left( \sqrt{ x^2 + \frac{D_s^2}{4} } - x \right) \right] - \cos \left[ k \left( \sqrt{ x^2 + \frac{D_s^2}{4} } - x \right) \right] \right) \\
&+ \frac{\alpha}{k_a} \left( 1 - \frac{x}{ \sqrt{ x^2 + \frac{D_s^2}{4} } } \right) \sin \left[ k_a \left( \sqrt{ x^2 + \frac{D_s^2}{4} } - x \right) \right] \Bigg) \, .
\end{split}
\end{equation}

\subsection{Normalization and dimensionless forms of the equations for the acoustic forcing field}

We shall finally now provide the dimensionless and normalized version of the equations we just derived.

Setting $\alpha = 0$ and $k_{\alpha} = \omega / c$ in Eqs.~\eqref{eq:analytical_pressure_acoustic_axis} and \eqref{eq:analytical_intensity_acoustic_axis} shows that $p_{ac}$ and $I_x$ are bounded respectively by ${p_{ac,max} = 2 \rho c u_0}$ and $I_{max} = 2 \rho c u_0^2$ in the unattenuated limit. Since $p_{ac} = \rho c u_{ac}$ for a plane wave, this also implies $u_{ac,max} = 2 u_0$. All these quantities can be related to the acoustic power $P_{ac}$ radiated by the transducer through:
\begin{equation} \label{eq:acoustic_power_definition}
P_{ac} = \iint_{S_t} \bm{I} \cdot \bm{e_x} \, dS = \frac{\rho c u_0^2}{8} \pi D_s^2 =  \frac{ I_{max} \pi D_s^2 }{16} \, ,
\end{equation}
in which the pressure at the transducer has been taken as $\rho c u_0$. Equations~\eqref{eq:Rayleigh_integral_all} are finally obtained (i) by normalizing Eqs.~\eqref{eq:acoustic_intensity_definition}, \eqref{eq:lossy_Rayleigh_integral_pressure} and \eqref{eq:lossy_Rayleigh_integral_velocity} by $I_{max}$, $p_{ac,max}$ and $u_{ac,max}$, respectively. (ii) The $\omega \tau \ll 1$ expressions of $k_{a}$ and $\alpha$ (Eq.~\eqref{eq:wavenumber_real_imaginary_parts}) are then used and (iii) made dimensionless using $D_s$ as length scale. A similar process applied to Eq.~\eqref{eq:analytical_intensity_acoustic_axis} yields the normalized on-axis intensity:
\begin{equation} \label{eq:analytical_intensity_acoustic_axis_dimensionless}
\begin{split}
\widetilde{I}_x (x, r=0) =& \frac{1}{4} e^{- \frac{N}{L} \left( \sqrt{x^2 + \frac{1}{4} } + x \right) } \Bigg( \left[ \frac{x}{ \sqrt{x^2 + \frac{1}{4} } } + 1 \right] \left( \cosh \left[ \frac{N}{L} \left( \sqrt{x^2 + \frac{1}{4} } - x \right) \right] - \cos \left[ \frac{2.44 \pi}{S} \left( \sqrt{ x^2 + \frac{1}{4} } - x \right) \right] \right) \\
&+ \left[ 1 - \frac{x}{ \sqrt{x^2 + \frac{1}{4} } } \right] \sinh \left[ \frac{N}{L} \left( \sqrt{ x^2 + \frac{1}{4} } - x \right) \right] + \frac{N S}{2.44 \, \pi L} \left[ 1 - \frac{x}{ \sqrt{x^2 + \frac{1}{4} } } \right] \\
&\times \sin \left[ \frac{2.44\pi}{S} \left( \sqrt{x^2 + \frac{1}{4} } - x \right) \right] \Bigg) \, ,
\end{split}
\end{equation}
which gives the far-field ($ 1 / \left( 4 x^2 \right) \ll 1$) approximation:
\begin{equation} \label{eq:analytical_intensity_axis_far_field}
\widetilde{I}_x(x, r=0) \sim \left( \frac{1.22 \pi}{8 \, S} \right)^2 \frac{ e^{-\frac{x}{L_{\alpha}} } }{x^2}\, .
\end{equation}
Equations~\eqref{eq:analytical_intensity_acoustic_axis_dimensionless} and \eqref{eq:analytical_intensity_axis_far_field} are respectively plotted as a solid and a dash-dotted lines in Fig.~\ref{fig:transducer_mesh_convergence} for the parameter values listed in Table~\ref{tab:dimensionless_parameters_values}.

\section{Grid sensitivity analysis} \label{appendix:grid_sensitivity_analysis}

\begin{figure}[h!]
	\centering
	\includegraphics[trim={0, 1.5cm, 0, 0}, clip, width=0.98\textwidth]{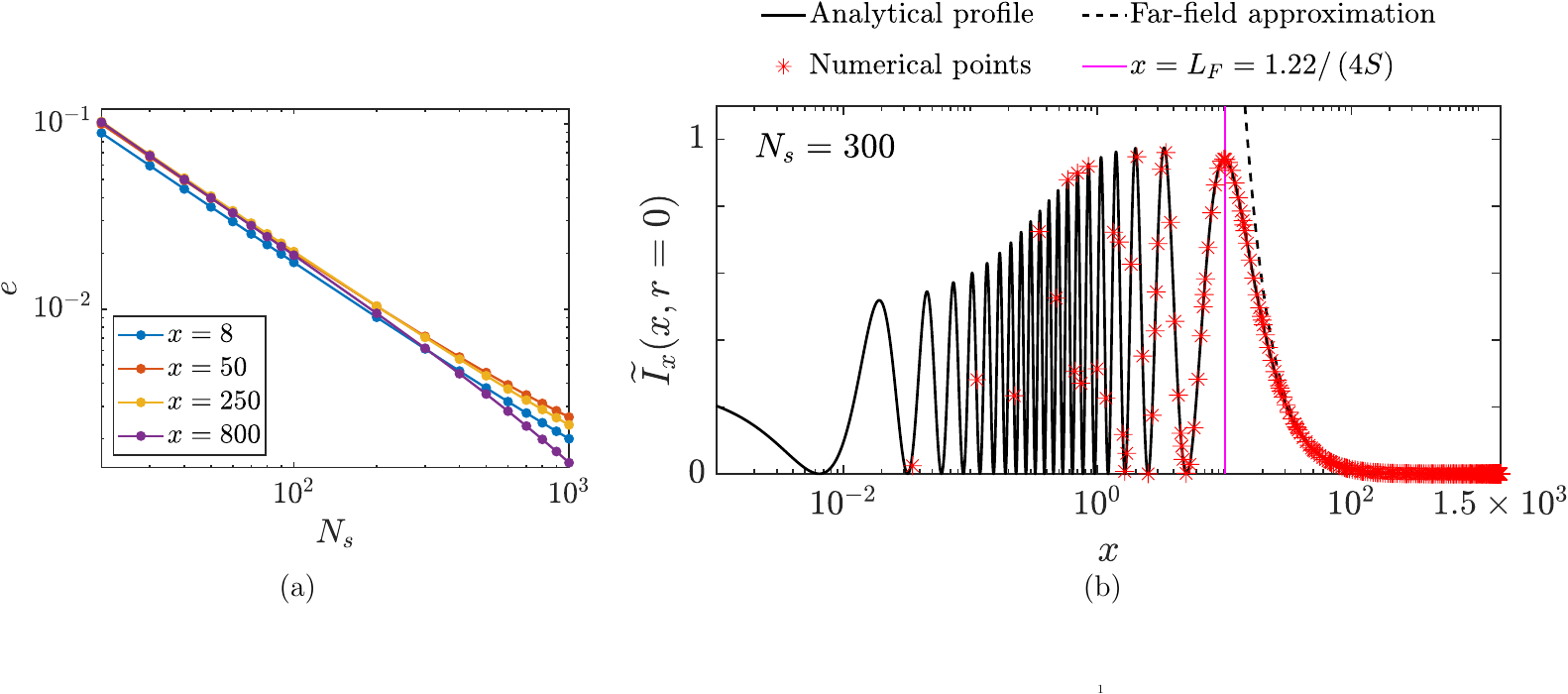}
	\caption{Effect of the transducer grid density on the acoustic force calculations. (a) Relative error $e$ between the numerical and analytical values of ${\widetilde{I}_{x}(x,r=0)}$ at different points on the axis, as a function of the number $N_s$ of point sources discretizing the transducer. (b) Comparison between the numerical (red, $N_s = 300$) and analytical (solid black curve, given by Eq.~\eqref{eq:analytical_intensity_acoustic_axis_dimensionless}) evaluations of ${\widetilde{I}_{x}(x,r=0)}$ over the domain length. The far-field approximation~\eqref{eq:analytical_intensity_axis_far_field} of $\widetilde{I}_x(x, r=0)$ (dashed black curve) is also displayed, and the vertical purple line locates the Fresnel distance $L_F$. In (b), the numerical computations of ${\widetilde{I}_{x}(x,r=0)}$ are made at the collocation points for $N_p=8$.}
	\label{fig:transducer_mesh_convergence}
\end{figure}

We validated the numerical computation of $\bm{\widetilde{I}}$ (and hence, the acoustic force field) and assessed the number $N_s$ of grid points on the transducer to be used for this purpose. The numerical evaluation of ${\widetilde{I}_{x}(x,r=0)}$ is compared to its analytical expression in Fig.~\ref{fig:transducer_mesh_convergence}~(a). The error on the numerically-calculated ${\widetilde{I}_{x}(x,r=0)}$ decreases rapidly to become lower than 1~\% for $N_s \geq 250$. Therefore, all intensity fields used in this work were computed with $N_s = 300$ to obtain an efficient balance between accuracy of the forcing field  calculations (Fig.~\ref{fig:transducer_mesh_convergence}~(b)) and computational cost.

The sensitivity to the grid discretizing the fluid domain has been assessed by carrying out $p$-refinement. The strong intensity gradients in the near acoustic field are undersampled to keep the numerical calculations tractable (Fig.~\ref{fig:transducer_mesh_convergence}~(b)). Thus, varying $N_p$ does not only act on the quality of the flow structures, it also affects the amount of momentum that is actually injected into the fluid domain.

Table~\ref{tab:mesh_convergence} shows the effect of $p$-refinement on the position $x_p$ and amplitude $u_{x,p}$ of the velocity peak on the jet axis, for the case of the highest forcing magnitude investigated. The errors on both $x_p$ and $u_{x,p}$ fall below $0.1$~\% for $N_p \geq 6$. Despite their non-monotonic evolution with $N_p$, caused by $\widetilde{\bm{I}}$ being undersampled in the acoustic near field, these errors remain extremely small for further-refined grids. The results shown in the reminder of this document were obtained for $N_p=8$, for which the effects of the acoustic near field undersampling on the flow are proven to be negligible.
\begin{table}[h]
    \caption{Dependence of the steady on-axis velocity peak position $x_p$ and amplitude $u_{x,p}$ with the polynomial degree $N_p$ of the expansion basis. The calculations are made for $Gr_{ac} = 5 \times 10^4$, i.e., for the case of highest forcing magnitude, and the errors are computed relatively to the results obtained for $N_p=9$. The values of $x_p$ are obtained by finding the point where $\partial u_x / \partial x$ vanishes on the axis, using the Lagrange polynomials of the expansion basis as interpolation functions.}
    \centering
    \begin{ruledtabular}
    \begin{tabular}{c c c c c}

        $N_p$ & $x_p$ & $u_{x,p}$ & Relative error on $x_p$ (\%) & Relative error on $u_{x,p}$ (\%)\\
        \hline
        $4$ & 23.187 & 736.05 & 2.2917 & 0.4584\\
        $5$ & 23.702 & 732.68 & 0.1256 & 0.0014\\
        $6$ & 23.748 & 732.45 & 0.0691 & 0.0334\\
        $7$ & 23.732 & 732.79 & 0.0012 & 0.0125\\
        $8$ & 23.729 & 732.76 & 0.0110 & 0.0084\\
        $9$ & 23.731 & 732.70 & $-$ & $-$
        
    \end{tabular}
    \end{ruledtabular}
    \label{tab:mesh_convergence}
\end{table}

\break
% Create the reference section using BibTeX:
\bibliography{references_with_DOI_abbreviated_journal_names.bib}
%\bibliography{references.bib}
\end{document}